\documentclass[twocolumn]{aastex631}

\usepackage{makecell}
\usepackage{pifont}
\newcommand{\cmark}{\ding{51}}%
\newcommand{\xmark}{\ding{55}}%

\defcitealias{Alam2022}{A22}

\usepackage{xcolor}

\usepackage{amsmath}

\newcommand{\TESS}{\textit{TESS\:}}
\newcommand{\CHEOPS}{\textit{CHEOPS\:}}

\newcommand{\bedit}[1]{#1}

\usepackage{etoolbox}
\makeatletter
\patchcmd{\ltx@foottext}{%
  .5\textwidth\advance\hsize-18pt}{%
  \linewidth\advance\hsize-1.8em%
}{}{}
\makeatother

\begin{document}

\title{A Ground-Based Transit Observation of the Long-Period Extremely Low-Density Planet HIP~41378~f}

\shorttitle{New HIP~41378~f Transit}

\author[0000-0003-1361-985X]{Juliana Garc\'ia-Mej\'ia}
\altaffiliation{51 Pegasi B Fellow, MIT Pappalardo Physics Fellow}
\affiliation{Kavli Institute for Astrophysics and Space Research, Massachusetts Institute of Technology, Cambridge, MA 02139, USA}
\affiliation{Center for Astrophysics \textbar\ Harvard \& Smithsonian, 60 Garden Street, Cambridge, MA 02138, USA}

\author[0000-0002-7564-6047]{Zo\"e L. de Beurs}
\altaffiliation{NSF Graduate Research Fellow, MIT Presidential Fellow, MIT Collamore-Rogers Fellow, MIT Teaching Development Fellow}
\affiliation{Department of Earth, Atmospheric and Planetary Sciences, Massachusetts Institute of Technology, Cambridge, MA 02139, USA}

\author[0000-0003-2171-5083]{Patrick Tamburo}
\affiliation{Center for Astrophysics \textbar\ Harvard \& Smithsonian, 60 Garden Street, Cambridge, MA 02138, USA}

\author[0000-0001-7246-5438]{Andrew Vanderburg}
\affiliation{Kavli Institute for Astrophysics and Space Research, Massachusetts Institute of Technology, Cambridge, MA 02139, USA}
\affiliation{Center for Astrophysics \textbar\ Harvard \& Smithsonian, 60 Garden Street, Cambridge, MA 02138, USA}

\author[0000-0002-9003-484X]{David Charbonneau}
\affiliation{Center for Astrophysics \textbar\ Harvard \& Smithsonian, 60 Garden Street, Cambridge, MA 02138, USA}

\author[0000-0001-6588-9574]{Karen A.\ Collins}
\affiliation{Center for Astrophysics \textbar\ Harvard \& Smithsonian, 60 Garden Street, Cambridge, MA 02138, USA}

\author[0000-0003-1464-9276]{Khalid Barkaoui}
\affiliation{Instituto de Astrof\'isica de Canarias (IAC), Calle V\'ia L\'actea s/n, 38200, La Laguna, Tenerife, Spain}
\affiliation{Astrobiology Research Unit, Universit\'e de Li\`ege, 19C All\'ee du 6 Ao\^ut, 4000 Li\`ege, Belgium}
\affiliation{Department of Earth, Atmospheric and Planetary Sciences, Massachusetts Institute of Technology, Cambridge, MA 02139, USA}

\author[0000-0001-8621-6731]{Cristilyn N.\ Watkins}
\affiliation{Center for Astrophysics \textbar\ Harvard \& Smithsonian, 60 Garden Street, Cambridge, MA 02138, USA}

\author[0000-0003-2163-1437]{Chris Stockdale}
\affiliation{Hazelwood Observatory, Australia}
 
\author[0000-0001-8227-1020]{Richard P. Schwarz} 
\affiliation{Center for Astrophysics \textbar\ Harvard \& Smithsonian, 60 Garden Street, Cambridge, MA 02138, USA}

\author[0000-0002-6482-2180]{Raquel For\'{e}s-Toribio}
\affiliation{Department of Astronomy, The Ohio State University, 140 West 18th Avenue, Columbus, OH 43210, USA}
\affiliation{Center for Cosmology and Astroparticle Physics, The Ohio State University, 191 W. Woodruff Avenue, Columbus, OH 43210, USA}

\author[0000-0001-9833-2959]{Jose A. Mu\~noz}
\affiliation{Departamento de Astronom\'{\i}a y Astrof\'{\i}sica, Universidad de Valencia, E-46100 Burjassot, Valencia, Spain}
\affiliation{Observatorio Astron\'omico, Universidad de Valencia, E-46980 Paterna, Valencia, Spain}

\author[0000-0002-8458-0588]{Giovanni Isopi}
\affiliation{Sapienza Università di Roma, Piazzale Aldo Moro, 5, 00185, Rome (RM), Italy}
\affiliation{Campo Catino Astronomical Observatory, Regione Lazio, Guarcino (FR), 03010, Italy}
\affiliation{INFN Sezione Roma1, Piazzale Aldo Moro, 2, 00185, Rome (RM), Italy}
\affiliation{INAF OAC, Via della Scienza, 5, 09047, Selargius (CA), Italy}

\author{Franco Mallia}
\affiliation{Campo Catino Astronomical Observatory, Regione Lazio, Guarcino (FR), 03010, Italy}

\author[0000-0002-9428-1573]{Aldo Zapparata}
\affiliation{Campo Catino Astronomical Observatory, Regione Lazio, Guarcino (FR), 03010, Italy}
\affiliation{Sapienza Università di Roma, Piazzale Aldo Moro, 5, 00185, Rome (RM), Italy}

\author[0000-0003-3184-5228]{Adam Popowicz}
\affiliation{Silesian University of Technology, Akademicka 16, Gliwice, Poland}

\author{Andrzej Brudny}
\affiliation{Silesian University of Technology, Akademicka 16, Gliwice, Poland}

\author[0000-0002-0802-9145]{Eric Agol}
\affiliation{Department of Astronomy, University of Washington, Box 351580, Seattle, WA 98195, USA}

\author[0000-0003-4157-832X]{Munazza K. Alam} 
\affiliation{Space Telescope Science Institute, 3700 San Martin Drive, Baltimore, MD 21218, USA}

\author[0000-0001-6285-9847]{Zouhair Benkhaldoun} 
\affiliation{Cadi Ayyad University, Oukaimeden Observatory, High Energy Physics, Astrophysics and Geoscience Laboratory, Faculty of Sciences Semlalia, Marrakech, Morocco}

\author[0000-0001-8923-488X]{Jehin Emmanuel} 
\affiliation{Space Sciences, Technologies and Astrophysics Research Institute, Universit\'e de Li\`ege, 19C All\'ee du 6 Ao\^ut, B-4000 Li\`ege, Belgium}

\author[0000-0003-3986-0297]{Mourad Ghachoui} 
\affiliation{Cadi Ayyad University, Oukaimeden Observatory, High Energy Physics, Astrophysics and Geoscience Laboratory, Faculty of Sciences Semlalia, Marrakech, Morocco}

\author[0000-0003-1462-7739]{Michaël Gillon} 
\affiliation{Astrobiology Research Unit, Universit\'e de Li\`ege, 19C All\'ee du 6 Ao\^ut, 4000 Li\`ege, Belgium}

\author[0000-0003-1728-0304]{Keith Horne}
\affiliation{SUPA Physics and Astronomy, University of St. Andrews, Fife, KY16 9SS Scotland, UK}

\author[0000-0003-0987-1593]{Enric Palle}
\affiliation{Instituto de Astrof\'isica de Canarias (IAC), Calle V\'ia L\'actea s/n, 38200, La Laguna, Tenerife, Spain}
\affiliation{Astrobiology Research Unit, Universit\'e de Li\`ege, 19C All\'ee du 6 Ao\^ut, 4000 Li\`ege, Belgium}

\author[0000-0003-3904-6754]{Ramotholo Sefako} 
\affiliation{South African Astronomical Observatory, P.O. Box 9, Observatory, Cape Town 7935, South Africa}

\author[0000-0002-1836-3120]{Avi Shporer}
\affiliation{Kavli Institute for Astrophysics and Space Research, Massachusetts Institute of Technology, Cambridge, MA 02139, USA}

\author{Mathilde Timmermans} 
\affiliation{School of Physics \& Astronomy, University of Birmingham, Edgbaston, Birmingham B15 2TT, United Kingdom}
\affiliation{Astrobiology Research Unit, Universit\'e de Li\`ege, 19C All\'ee du 6 Ao\^ut, 4000 Li\`ege, Belgium}

\correspondingauthor{Juliana Garc\'ia-Mej\'ia}
\email{jgarciam@mit.edu}

\begin{abstract}

We present a ground-based transit detection of HIP~41378~f, a long-period (\(P = 542\)~days), extremely low-density (\(0.09 \pm 0.02\)~g\,cm\(^{-3}\)) giant exoplanet in a dynamically complex system. Using photometry from \textit{Tierras}, TRAPPIST-North, and multiple LCOGT sites, we constrain the transit center time to \(T_{C,6} = 2460438.891 \pm 0.052\)~BJD~TDB. This marks only the second ground-based detection of HIP~41378~f, currently the longest-period and longest-duration transiting exoplanet observed from the ground. We use this new detection, \bedit{along with a recently published transit time from Rossiter--McLaughlin observations,} to update the TTV solution for HIP~41378~f. \bedit{We predict the next two transits will occur at $T_{C,7} = 2460980.793^{+0.098}_{-0.129}$~BJD~TDB (2025 November 1) and $T_{C,8} = 2461522.653^{+0.213}_{-0.238}$~BJD~TDB (2027 April 27).} Incorporating new \TESS\ Sector~88 data, we also rule out the 101-day orbital period alias for HIP~41378~d, and find that the remaining viable solutions are centered on the 278, 371, and 1113-day aliases. The latter two imply dynamical configurations that challenge the canonical view of planet~e as the dominant perturber of planet~f. Our results suggest that HIP~41378~d may instead play the leading role in shaping the TTV of HIP~41378~f.

\end{abstract}

\section{Introduction}\label{sec:intro}

The planetary system around the bright ($V=8.93$) F-type star HIP~41378 is a unique dynamical and chemical laboratory. First identified by \citet{Vanderburg2016}, the system includes at least five transiting planets with orbital periods ranging from 15 days to approximately 1.5 years \citep{Becker2019,Berardo2019,Santerne2019}, occupying a regime between the cold giants of the outer Solar System and the highly irradiated hot Jupiters. As a result, these planets are particularly interesting targets for dynamical and comparative atmospheric studies.

Despite the breadth of observational campaigns that have targeted the system, many aspects of its architecture remain uncertain. The two inner planets, HIP~41378~b and c, are well characterized with short periods ($P_b = 15.5712$ d, $P_c = 31.6978$ d) and repeated transits. However, the three outer planets, HIP~41378~d, e, and f, pose greater challenges due to their long orbital periods \bedit{($P > 100$ d)} and limited transit detections. Planet d has been observed to transit twice (in \textit{K2} C5 and C18), and planet e only once (during \textit{K2} C5), leaving their orbital periods and dynamical roles poorly constrained. 

\citet{Becker2019} and \citet{Berardo2019} derived 23 possible orbital periods for planet d based on its two observed transits. Subsequent analyses have narrowed this list: \citet{Grouffal2022} used \TESS\ data to eliminate 16 candidate solutions and suggested a tentative Rossiter–McLaughlin (RM) detection consistent with a 278-day period. However, \citet{Sulis2024} recently reported a non-detection of a transit in that period with \CHEOPS\, indicating either a different orbital period or the presence of significant transit timing variations (TTVs). \citet{Sulis2024} narrowed down the possible periods for planet d to only four options: 101, 278, 371, and 1113 days. The period of planet e remains elusive, with only a loose constraint of $P_e = 260^{+160}_{-60}$ days based on the transit shape and stellar parameters \citep{Lund2019}\bedit{, and an RV-derived estimate of $369\pm10$ days \citep{Santerne2019} (though the latter assumed planet d's period was fixed at 278 days).}

HIP~41378~f, the largest known planet in the system, \bedit{is} the best-characterized member of the outer planet trio. Detected in two \textit{K2} transits and followed up via a multi-year radial velocity (RV) campaign \citep{Santerne2019}, a ground-based transit \citep{Bryant2021}, an \textit{HST/WFC3} transmission spectrum (\citealt{Alam2022}, hereafter \citetalias{Alam2022}), \bedit{and a recent RM campaign \citep{Grouffal2025}}, its period is currently constrained at \bedit{$542.0797 ^{+0.0001}_{-0.0002}$ days, though TTVs of several hours have been measured}. HIP~41378~f has a radius of \( R_p = 9.2 \pm 0.1 \, R_{\oplus} \) and a mass of \( M_p = 12 \pm 3 \, M_{\oplus} \), implying an extraordinarily low bulk density (\(0.09 \pm 0.02\) g cm\(^{-3}\)). This extreme value defies traditional planet formation models \citep[e.g.,][]{Mordasini2012,Belkovski2022}, and has prompted the proposal of alternative explanations, such as high-altitude photochemical hazes \citep{Gao2020,Ohno2021} or the presence of rings \citep{Zuluaga2015, Akinsanmi2020,Piro2020}. The flat transmission spectrum of the planet obtained with HST is consistent with both scenarios and remains inconclusive (\citetalias{Alam2022}).

A promising avenue to understand the low density of HIP~41378~f is atmospheric characterization using facilities such as JWST. Owing to its long orbital period and a known TTV signal \citep{Bryant2021}, precise transit predictions are crucial to schedule such observations. Previous studies (\citealt{Bryant2021} and \citetalias{Alam2022}) have assumed that HIP~41378~e is more massive and on a wider orbit than planet d \citep{Santerne2019}, and thus the dominant source of the TTVs observed for planet~f. This assumption is based on the expected proximity of planets e and f to a 3:2 mean-motion resonance (MMR) and the scaling of the TTV amplitude with the perturber's mass and period \citep{Agol2005,Holman2005}. However, because of the poorly constrained orbits of both planets d and e, it remains unclear which is the true source of the perturbations. If planet d, for example, lies closer to an MMR with planet~f than currently assumed, it could instead be the dominant contributor to the TTV signal. 

In this paper, we present a new measurement of the 2024 May transit of HIP~41378~f from ground-based photometry and evaluate its implications for the system’s TTV signal. Section~\ref{sec:data} details our ground-based observations. \bedit{In Section~\ref{sec:photometry}, we describe our photometric extraction. In Section \ref{sec:detection} we inspect the light curves and identify flux decrements consistent with the transit depth of HIP~41378~f. In Section~\ref{sec:analysis}, we model the light curve and constrain the transit center time (\S \ref{sec:lcfit}), use \TESS\ observations to update the possible orbital solutions for planet d (\S \ref{sec:prettv}), and present a revised fit to the TTV signal of HIP~41378~f (\S \ref{sec:ttvfit}).} We summarize our work in Section~\ref{sec:discussion}.

\section{Observations}\label{sec:data}

The transit duration of HIP~41378~f is approximately 19 hours, making it impossible to observe a full transit from any single Earth-based observatory. Therefore, we initiated a multi-telescope campaign with two observational goals: first, to detect ingress or egress; and second, to employ precise night-to-night photometers to measure the transit signal even if ingress or egress was not directly observed. We leveraged the \TESS\-Exoplanet Follow-up Observing Program Sub-Group 1 (TFOP SG1, Collins 2019) to observe HIP~41378 during the transit window predicted by \citetalias{Alam2022}. \bedit{Specifically, we requested two-hour (or longer) photometric sequences from all participating sites between UTC 2024 May 7 18:00 and UTC 2024 May 9 04:00, corresponding to the $\pm1\sigma$ window from \citetalias{Alam2022}'s prediction ($T_{C,6} = 2460438.95 \pm 0.02$ BJD TDB) with additional margin to account for potential TTVs of several hours based on the observed TTV signal.}

\subsection{\textit{Tierras}}\label{sec:datatierras}

The \textit{Tierras} Observatory is a refurbished 1.3~m ultra-precise fully-automated photometer located at the F.L. Whipple Observatory atop Mount Hopkins, Arizona \citep{GaMe20}. \textit{Tierras} uses a custom narrow bandpass filter ($\lambda_C = 863.5$ nm, FWHM $=40$ nm) to minimize precipitable water vapor (PWV) errors that are known to limit photometric precision from the ground \citep[e.g.,][]{St07,Bl08,Be12}. \textit{Tierras} employs a $4K \times 4K$ e2v CCD ($15~\mu$m pixel size) with a plate scale of \bedit{$0\farcs43/$pixel} operated in frame-transfer mode, yielding a field of view (FOV) of $15' \times 29'$ (R.A.$\times$decl.). The facility is currently achieving night-to-night precisions as low as 0.5~ppt in months-long baselines for bright, non-variable targets \citep[e.g.,][]{Tamburo2025}. 

We observed HIP~41378~f with \textit{Tierras} on UT~2024~May~7, May~8, and May~9. The three nights were clear, with FWHM seeing values averaging \bedit{$1\farcs8$, $1\farcs6$, and $1\farcs7$}, respectively. \bedit{We bias-corrected the \textit{Tierras} images using the facility pipeline, which subtracts overscan-derived bias levels independently for each amplifier and subsequently stitches the detector halves. We did not flat-field the data because of additive glow from off-axis scattering within the optical path, which has been addressed since the observations were taken with additional optical baffling within the instrument. To compensate for not flat fielding the data, we restricted our analysis to images with pointing deviations of less than 2.5~pixels (1.1\arcsec) in either the $x$ or $y$ directions. This threshold selects images that are positioned to within one average seeing FWHM on each night, ensuring that approximately the same pixels are sampled in each exposure and minimizing uncertainties introduced by non-uniform pixel sensitivities. This process removed 19 images.}

\subsection{TRAPPIST-North}

The TRAnsiting Planets and PlanetesImals Small Telescope in the North (TRAPPIST-North) is a 60-cm $f/8$ Ritchey-Chr\'etien robotic telescope at Oukaimeden Observatory, in Morocco \citep{Barkaoui2019_TN}. It is a twin of TRAPPIST-South, located at La Silla Observatory, in Chile \citep{Gi11,Je11}. The TRAPPIST-North telescope is equipped with an Andor iKon-L BEX2-DD deep-depletion $2K \times 2K$ e2v CCD ($15~\mu$m pixel size), yielding a FOV of $20' \times 20'$ and a plate scale of \bedit{$0\farcs60 /$pixel}. 

We observed HIP~41378~f with TRAPPIST-North on UT~2024~May~7 and May~8 through the $B$ filter with an exposure time of 20~s. Observation conditions were clear on both nights, with low humidity and wind. \bedit{We performed initial data calibration (flat-fielding and bias correction) using the {\tt PROSE} pipeline \citep{prose}.}

\subsection{LCOGT}
We gathered observations with 1.0\,m  and 0.35\,m network nodes of the Las Cumbres Observatory Global Telescope \citep[LCOGT;][]{Br13}. The 1.0\,m nodes we used are located at Teide Observatory on the island of Tenerife (TEID), Cerro Tololo Inter-American Observatory in Chile (CTIO), and McDonald Observatory near Fort Davis, Texas, United States (McD). The 0.35\,m network nodes we used are located at TEID, CTIO, McD, and Haleakala Observatory on Maui, Hawai'i (HAl). The 1.0\,m telescopes are equipped with \bedit{$4K\times4K$} SINISTRO cameras that have an image scale of \bedit{$0\farcs39/$pixel}, resulting in a $26'\times26'$ FOV. The 0.35\,m Delta Rho 350 Planewave telescopes are equipped with QHY600 CMOS cameras with \bedit{$9.6K\times6.4K$} pixels and an image scale of $0\farcs73$ per pixel, resulting in a FOV of \bedit{$1\fdg9\times1\fdg2$. For this campaign, the 0.35\,m telescopes were operated in a cropped readout mode that utilized only the central $2.4K \times 2.4K$ pixels ($30'\times30'$) to reduce data volume and readout time.}

We observed HIP~41378~f using six 0.35\,m and five 1.0\,m telescopes across the LCOGT network. Details of the observing sequences, including the facility, date, exposure time, and filter used, are provided in Table~\ref{tab:obs}. \bedit{We flat-fielded and bias-corrected all the LCOGT images using the standard {\tt BANZAI} pipeline \citep{McCully:2018}.}

\subsection{Hazelwood Observatory}

The Hazelwood Observatory (Hwd) is located in Victoria, Australia, and is a private backyard observatory with a 0.32\,m Planewave CDK $f/8$ telescope equipped with an SBIG STT3200 $2.2K \times 1.5K$ KAF-3200 CCD cooled to $-40^{\circ}$C. The facility has a FOV of $20\arcmin \times13\arcmin $ and a plate scale of \bedit{$0 \farcs 55 / $pixel}. 

\bedit{Hwd attempted to observe a possible early transit of HIP~41378~f, taking 291 exposures of 20 sec through the Sloan $i'$ filter on UTC 2024 May 7. Skies were clear during these observations.}

\subsection{Observatori Astronòmic de la Universitat de València}

Observatori Astron\`omic de la Universitat de Val\`encia (OAUV) hosts a 0.5\,m telescope in Aras de los Olmos, Valencia, Spain. The telescope is equipped with a $4K\times4K$ FLI ProLine camera, 9~$\mu$m pixels, \bedit{a FOV of $37' \times 37'$,} and a plate scale of $0\farcs54/$pixel. \bedit{The night of the observation, UTC 2024 May 7, was dark and, given the brightness of the star, the telescope was out of focus to avoid target saturation.}

\bedit{We used OAUV to gather 328 10-sec images of HIP~41378~f through the $R$ filter spanning 157 minutes.}

\subsection{Campo Catino Astronomical Observatory}

The Campo Catino Astronomical Observatory (OACC) remotely operates the Campocatino Austral Observatory (CAO), located at El Sauce Observatory in Río Hurtado, Chile. CAO uses a 0.6\,m $f/6.5$ PlaneWave CDK24 Corrected Dall-Kirkham telescope equipped with a \bedit{$9.6K\times6.4K$} Sony IMX455 CMOS detector with $3.76\mu$m pixels, which yields a plate scale of \bedit{$0\farcs2/$pixel and a FOV of $32' \times 21'$.}

\bedit{We observed HIP~41378~f with OACC-CAO on UTC 2024 May 9 under clear skies and favorable conditions. We gathered 146 35-sec exposures through the Sloan $i'$ filter.}

\subsection{Silesian University of Technology Observatory}

Silesian University of Technology Observatory (SUTO) is located in Otívar, Spain. It consists of a 0.3\,m Ritchey–Chrétien telescope mounted on a Paramount ME and housed inside a dome. The telescope is equipped with a cooled ASI ZWO1600MM CMOS camera with \bedit{$4.7K\times3.5K$} pixels of size $3.8~\mu$m, providing an image scale of \bedit{$0\farcs68/$pixel and a FOV of $53'\times40'$.} 

\bedit{We observed HIP~41378~f with SUTO on UTC 2024 May 8 through the $R$ filter with an exposure time of 15 s (237 exposures in total). The conditions were good and stable throughout the night.}

\begin{deluxetable*}{lcccccc}
\label{tab:obs}
\setlength{\tabcolsep}{3pt} 
\tabletypesize{\scriptsize}
\tablecaption{HIP~41378 observations sorted by facility, telescope designation (T1 or T2), and observation date. The table includes the number of images, exposure time, airmass range, and elevation range. \label{tab:hip41378_observations}}
\tablehead{
\colhead{Facility} & 
\colhead{UTC Date\tablenotemark{\scriptsize a}} & 
\colhead{\# of Images} & 
\colhead{\makecell{Exposure \\ Time (s)}} & 
\colhead{Filter} &
\colhead{\makecell{Airmass \\ Range}} & 
\colhead{\makecell{Elevation (°) \\ Range}} 
}
\startdata
\textit{Tierras} & 2024-05-07 & 210 & 8 & Custom & 1.66--2.35 & 37--25 \\
\textit{Tierras}\tablenotemark{\scriptsize e} & 2024-05-08 & 351 & 8 & Custom & 1.30--1.86 & 50--33 \\
\textit{Tierras} & 2024-05-09 & 404 & 8 & Custom & 1.3\bedit{7}--2.29 & 47--26 \\
TRAPPIST-North & 2024-05-07 & 220 & 20  & \textit{B} & \bedit{1.36--2.99} & 50--13 \\
TRAPPIST-North & 2024-05-08 & 163 & 20  & \textit{B} &  \bedit{1.38}--2.92 & 52--20 \\
LCOGT TEID 0.35\,m\tablenotemark{\scriptsize b} & 2024-05-07 & 299 & 20 & Sloan $i'$ & 1.24--2.32 & 54--26 \\
LCOGT TEID 1.0\,m & 2024-05-07 & 62 & 10 & Sloan~$z_s$& 1.24--1.42 & 45-54 \\
LCOGT HAl 0.35\,m, T1 & 2024-05-08 & 563 & 8 & Sloan $i'$  & 1.16--2.13 & 60--28 \\
LCOGT HAl 0.35\,m, T2 & 2024-05-08 & 551 & 8 & Sloan $i'$  & 1.16--2.10 & 60--28 \\
LCOGT McD 0.35\,m\tablenotemark{\scriptsize e} & 2024-05-08 & 456 & 8 & Sloan $i'$ & 1.28--\bedit{2.38} & 52--24 \\
LCOGT McD 0.35\,m & 2024-05-09 & 519 & 8 & Sloan $i'$ & 1.29--2.41 & 51--25 \\
LCOGT McD 1.0\,m, T1 & 2024-05-08 & 201 & 10 & Sloan~$z_s$
 & 1.28--2.47 & 52--24 \\
LCOGT McD 1.0\,m, T1 & 2024-05-09 & 196 & 10 & Sloan~$z_s$ & 1.29--2.47 & 51--24\\
LCOGT McD 1.0\,m, T2\tablenotemark{\scriptsize c} & 2024-05-08 & 53 & 10 & Sloan~$z_s$ & 1.28--1.43 & 52--44 \\
LCOGT McD 1.0\,m, T2 & 2024-05-09 & 188 & 10 & Sloan~$z_s$ & 1.29--2.46 & 51--24 \\
LCOGT CTIO 0.35\,m, T1 & 2024-05-09 & 743 & 8 & Sloan $i'$ & 1.35--2.81 & 48--21 \\
LCOGT CTIO 0.35\,m, T2 & 2024-05-09 & 745 & 8 & Sloan $i'$ & 1.35--2.82 & 48--21 \\
LCOGT CTIO 1.0\,m, T1\tablenotemark{\scriptsize d}  & 2024-05-09 & 274 & 10 & Sloan~$z_s$ & 1.35--2.81 & 48--21 \\
LCOGT CTIO 1.0\,m, T2 & 2024-05-09 & 273 & 10 & Sloan~$z_s$ & 1.35--2.80 & 48--21 \\
Hwd 0.32\,m & 2024-05-07 & 290 & 20 & Sloan $i'$ & 1.57--2.62& 40--22\\
OAUV 0.50\,m & 2024-05-08 & 328 & 10 & Johnson $R$ & 1.43--3.76 & 45--16 \\
OACC-CAO 0.60\,m & 2024-05-09 & 146 & 35 & Sloan $i'$ & 1.44--2.73 & 44--21 \\ 
SUTO-Otivar 0.3\,m & 2024-05-09 & 237 & 15 & Johnson $R$ & 1.43--3.57 & 44--16\\ 
\enddata
\tablenotetext{a}{The UTC Date listed is the start date of the observing sequence.} 
\vspace{-7pt}
\tablenotetext{b}{Data from the LCOGT TEID 0.35\,m telescope were saturated and are therefore excluded from the remainder of this work.}
\vspace{-7pt}
\tablenotetext{c}{The observing sequence for the LCOGT McD 1.0\,m, T2 telescope was accidentally interrupted by the observing robot \bedit{(ending at BJD TDB 2460438.633, approximately 1.6 hours into the night).}}
\vspace{-7pt}
\tablenotetext{d}{Observations from the LCOGT CTIO 1.0\,m, T1 telescope were obstructed by the dome and are consequently omitted from this work.}
\vspace{-14pt}
\bedit{\tablenotetext{e}{The transit dips used to constrain $T_{C,6}$ were detected on UTC 2024-05-08 in the \textit{Tierras} and LCOGT McD 0.35\,m datasets.}}
\end{deluxetable*}

\section{\bedit{Photometric Extraction}\label{sec:photometry}}

\bedit{We extracted photometry from all datasets gathered during the campaign. Given the diversity of telescopes, observing cadences, and baseline lengths across our observations, we employed two distinct photometric extraction approaches optimized for each dataset type.}

\bedit{\subsection{\textit{Tierras}, TRAPPIST-North, and LCOGT}\label{sec:multinight}}

\bedit{For datasets spanning multiple nights (\textit{Tierras}, TRAPPIST-North, and all LCOGT facilities), we performed uniform photometric re-extraction optimized for detecting the transit through night-to-night flux comparisons. To minimize systematic errors introduced by differences in analysis pipelines, we used a consistent reduction approach based on the \textit{Tierras} pipeline. This ensures that differences between facilities reflect genuine astrophysical or instrumental effects rather than artifacts of heterogeneous data processing.}

\bedit{The pipeline uses the World Coordinate System (WCS) of the images to identify all sources within the detector footprint up to a magnitude limit of Gaia $G_\mathrm{RP} = 17$~mag, and places aperture radii ranging from $5$ to $25$~pixels, as well as circular annuli (inner radius of $25$~pix, outer radius of $35$~pix) to measure the local background of each source. The background is taken to be the median of the sigma-clipped pixels within the annulus with a clipping level of 2$\sigma$. We calculate the expected uncertainty on the target's normalized photometry including contributions from photon noise (target + sky), dark current, and read noise, with the former being the dominant source of error.}

\bedit{We then construct an ensemble light curve (ELC) by computing a weighted sum of the fluxes from all comparison stars. Following a method based on \citet{Tamburo2022}, we assign weights to each star through an iterative convergence procedure that optimizes photometric precision while disfavoring noisy stars. This algorithm proceeds as follows: stars are initially weighted by their computed photometric errors, and each star's flux is corrected by the ELC constructed from all other stars. The weights are then updated using the \textit{measured} standard deviation of each corrected light curve. This procedure is repeated until convergence.}

\bedit{We first perform a ``crude" convergence loop requiring normalized weights to stabilize within $1e-3$ of their previous values. We then compute the ratio of measured to expected standard deviation for each light curve and perform sigma clipping on these ratios, retaining only stars with ratios below the 1-$\sigma$ upper limit as potential references. This removes the noisiest stars from contributing to the ELC. Finally, we perform a ``fine" convergence loop with the remaining stars, requiring weights to converge within $1e-6$ of their previous values.}

\bedit{Critically, once converged, these weights remain fixed and are applied uniformly across all nights of observation for each telescope. This ensures that the ELC construction is consistent across nights and that any observed flux variations reflect genuine astrophysical or instrumental differences rather than artifacts from varying comparison star contributions. The target flux is divided by the ELC and normalized. An estimate of scintillation noise \citep[e.g.,][]{Stefansson2017} is added to the propagated uncertainties from the ELC correction. The final light curve uses the aperture size that yields the lowest measured scatter for the target.}

\begin{figure}[h!]
    \centering
    \includegraphics[width=\linewidth]{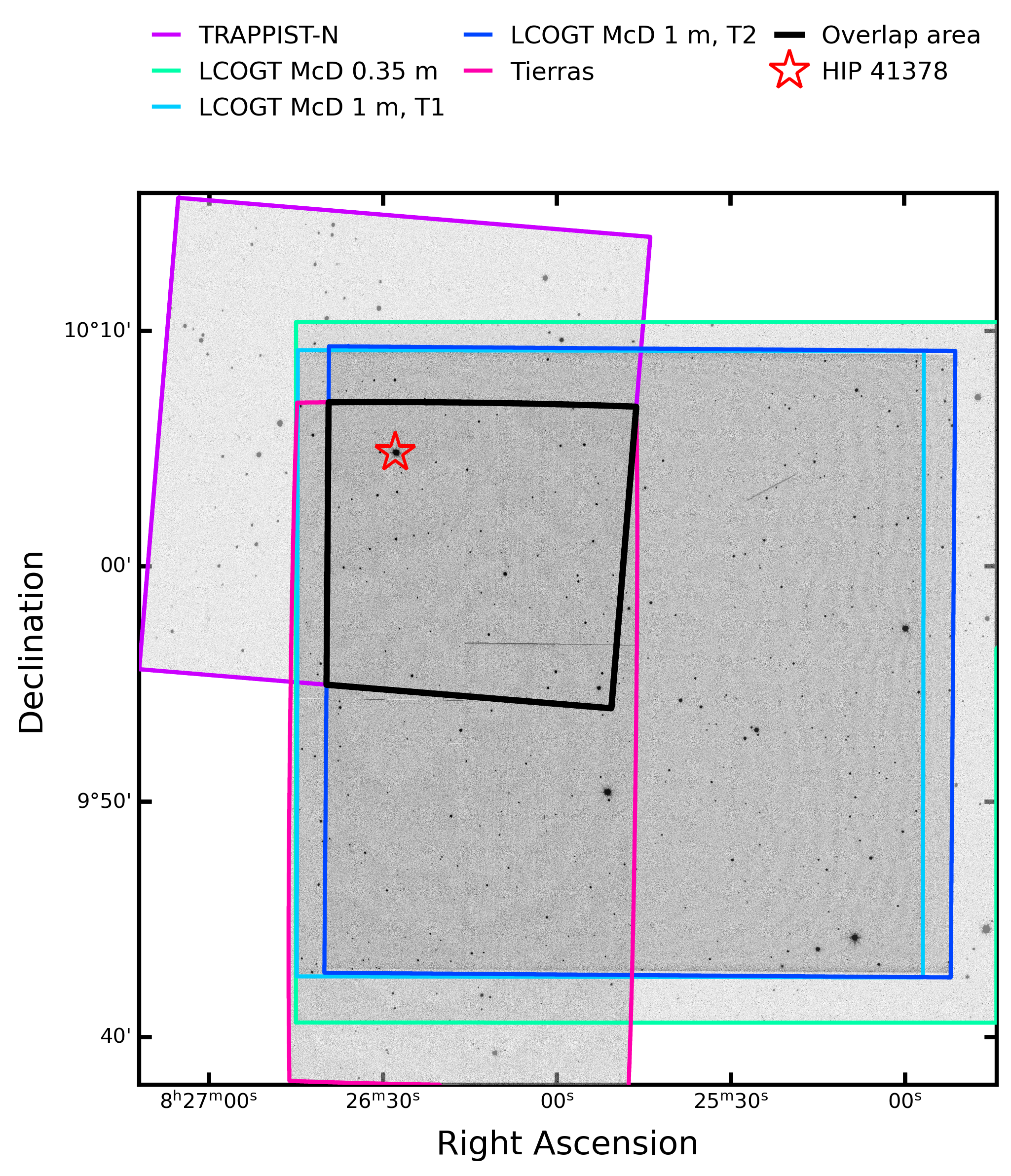} 
    \bedit{\caption{Sky-projected field of view of the observatories from where multi-night data sets were gathered, in Right Ascension and Declination coordinates. The field outlines are color-coded as follows: \textit{Tierras} is shown in pink, TRAPPIST-North in violet, LCOGT McD 0.35\,m in teal, McDonald 1.0\,m (T1) in light blue, and McDonald 1.0\,m (T2) in dark blue. The thick black outline shows the flux-limited overlapping field of view between facilities, which contains only $\sim$6\% of HIP~41378's flux in available comparison stars. HIP~41378 is highlighted with a red star symbol.}}
    \label{fig:FOVs}
\end{figure}

\bedit{A critical question arose while building the ELC: why not use a common set of comparison stars for \textit{Tierras}, TRAPPIST-North, and LCOGT? As illustrated in Figure~\ref{fig:FOVs}, stars within the flux-limited overlap region (black outline) provide only about $6\%$ of HIP~41378's flux. For reliable differential photometry, the comparison star flux should substantially exceed the target flux to minimize systematic errors \citep{Nutzman2008}. With such a low flux ratio, photometric noise from the ensemble comparison would dominate, precluding reliable detection of the $\sim$4.4\,ppt transit signal of HIP~41378~f.}

\bedit{Given these constraints, we instead optimized comparison star selection independently for each facility to maximize photometric precision using the ELC method described above. While the specific comparison star sets differ between facilities, the same set of comparison stars and their converged ELC weights were used consistently for all nights of observation from each individual telescope. Table~\ref{tab:photometry} summarizes the aperture radii, number of comparison stars, and 2\,min RMS achieved for \textit{Tierras}, TRAPPIST-North, and all LCOGT telescopes.}

\bedit{An additional consideration emerged regarding the LCOGT network data. In principle, the LCOGT network's design philosophy (identical telescope classes distributed globally) could enable treating all 0.35\,m telescopes as a single instrument and all 1.0\,m telescopes as another, allowing out-of-transit baselines from one site to normalize in-transit data from another site within the same telescope class. However, we adopted a more conservative approach for this analysis. Night-to-night photometric stability at the sub-mmag level (required to detect HIP~41378~f's shallow transit) is known to be significantly affected by variations in PWV content in Earth's atmosphere \citep[e.g.,][]{St07,Bl08,Be12}. Water vapor can vary substantially not only between geographically distinct sites but also at the same location from night to night \citep{Baker2017}, even when using identical instrumentation. Given these considerations, we define a multi-night facility as one where the same individual telescope observed on two or more nights, rather than grouping all telescopes of the same class together. \bedit{Under this definition, only the \textit{Tierras}, TRAPPIST-North, and LCOGT McD telescopes qualify as multi-night facilities.} Nevertheless, we applied our uniform photometric extraction procedure (described above) to all LCOGT datasets to ensure internal consistency in our photometric analysis and to facilitate potential future work.} 

\bedit{The photometric stability assessment described in Section~\ref{sec:ntn_stable} demonstrates that both McD 1.0\,m telescopes fail to meet the requirements for inclusion in our transit fit. \bedit{Figure~\ref{fig:lcfit} therefore only shows the extracted light curves for the three multi-night facilities used in the transit fit: \textit{Tierras}, TRAPPIST-North, and LCOGT McD 0.35\,m.}}

\begin{deluxetable}{lrrrr}
\label{tab:photometry}
\tabletypesize{\scriptsize}
\tablecaption{Aperture radii, number of reference stars, and measured standard deviations of light curves from \bedit{\textit{Tierras}, TRAPPIST-North, and all LCOGT telescopes, reduced with the \textit{Tierras} pipeline. The reported standard deviations were evaluated using data binned to 2 and 10\,min cadences. Since the observed transit would bias the measured standard deviations upward for facilities that observed both in- and out-of-transit data, we first median-normalized each individual night before assessing $\sigma_\mathrm{2\,min}$ and $\sigma_\mathrm{10\,min}$ so that the precision of the datasets can be compared directly.}}
\tablehead{
    \colhead{Facility} & 
    \colhead{\makecell{Aperture \\ Rad. (pix)}} & 
    \colhead{\makecell{\# of Ref. \\ Stars}} & 
    \colhead{\makecell{$\sigma_\mathrm{2\,min}$ \\ (ppt)}} &
    \bedit{\colhead{\makecell{$\sigma_\mathrm{10\,min}$ \\ (ppt)}}}
}
\startdata
\textit{Tierras} & 12 & 88 & 1.3 & 0.8 \\
TRAPPIST-North & 21 & 73 & 2.8 & 1.7 \\
LCOGT TEID 1.0\,m & 12 & 129 & 1.3 & 0.5 \\
LCOGT HAl 0.35\,m, T1 & 11 & 162 & 1.9 & 0.9 \\
LCOGT HAl 0.35\,m, T2 & 13 & 171 & 2.2 & 0.8 \\
LCOGT McD 0.35\,m & 9 & 163 & 3.5 & 1.4 \\
LCOGT McD 1.0\,m, T1 & 12 & 127 & 2.6 & 1.3 \\
LCOGT McD 1.0\,m, T2 & 9 & 130 & 2.5 & 1.5 \\
LCOGT CTIO 0.35\,m, T1 & 11 & 169 & 2.2 & 1.0 \\
LCOGT CTIO 0.35\,m, T2 & 21 & 174 & 1.8 & 0.8 \\
LCOGT CTIO 1.0\,m, T2 & 25 & 129 & 1.7 & 0.8 \\
\enddata
\end{deluxetable}

\subsection{\bedit{Hwd, OAUV, OACC-CAO, and SUTO}}\label{sec:single_phot}

\bedit{For single-night datasets from facilities outside the LCOGT network (Hazelwood Observatory, OAUV, OACC-CAO, and SUTO), light curves were reduced using facility-standard pipelines and software packages. Image calibration procedures varied by facility: Hwd used MaxImDL v.6.50 for bias correction and flat-fielding, while OAUV, OACC-CAO, and SUTO used AstroImageJ \citep[AIJ;][]{Collins2017}. All four facilities used AIJ for photometric extraction and airmass detrending. \bedit{Table~\ref{tab:photometry_singlenight} summarizes the photometric extraction parameters and achieved precisions for these facilities.} The light curves from Hwd, OAUV, OACC-CAO, and SUTO are shown in Figure \ref{fig:singlenight_lcs} along with the single-night light curves from LCOGT TEID, CTIO, and Hal}. 

\begin{deluxetable}{lrrrr}
\label{tab:photometry_singlenight}
\tabletypesize{\scriptsize}
\tablecaption{\bedit{Aperture radii, number of reference stars, and measured standard deviations of light curves from single-night non-LCOGT facilities, reduced with AstroImageJ. The reported standard deviations were evaluated using data binned to 2 and 10\,min cadences. Note that these facilities observed only a single night and were not used in the transit center fit.}}
\tablehead{
    \colhead{Facility} & 
    \colhead{\makecell{Aperture \\ Rad. (pix)}} & 
    \colhead{\makecell{\# of Ref. \\ Stars}} & 
    \colhead{\makecell{$\sigma_\mathrm{2\,min}$ \\ (ppt)}} &
    \bedit{\colhead{\makecell{$\sigma_\mathrm{10\,min}$ \\ (ppt)}}}
}
\startdata
Hwd 0.32\,m & 9 & 6 & 3.5 & 1.9 \\
OAUV 0.50\,m & 19 & 12 & 5.6 & 3.0 \\
OACC-CAO 0.60\,m & 30 & 20 & 1.5 & 0.8 \\
SUTO-Otivar 0.30\,m & 17 & 17 & 3.9 & 2.4 \\
\enddata
\end{deluxetable}

\section{Did we detect the transit of HIP~41378~\MakeLowercase{f}?}\label{sec:detection}

\subsection{\bedit{Initial Light Curve Assessment}}

\bedit{We inspected the light curves from each facility by eye and found them to be flat, showing 
no intra-night flux variations consistent with the $\sim$72-minute ingress or egress of 
HIP~41378~f \citep{Vanderburg2016}. To quantify this assessment, we evaluated the photometric 
precision of each dataset in 10-minute bins (Tables~\ref{tab:photometry} and 
\ref{tab:photometry_singlenight}). We calculated the signal-to-noise ratio (SNR) for detecting 
a 72-minute, 4.4~ppt flux decrement as SNR~=~4.4~ppt~/($\sigma_{10\,\mathrm{min}}/\sqrt{7.2}$). 
All facilities achieve detection sensitivities $\geq$3.9$\sigma$, with \textit{Tierras} 
reaching up to 17.7$\sigma$. Visual inspection confirmed the absence of any flux decrements 
at these significance levels. We therefore conclude that none of our observations captured 
the planet during ingress or egress and turn to multi-night datasets to search for the 
transit through night-to-night photometric comparisons.}

\begin{figure*}
    \centering
    \includegraphics[width=\linewidth]{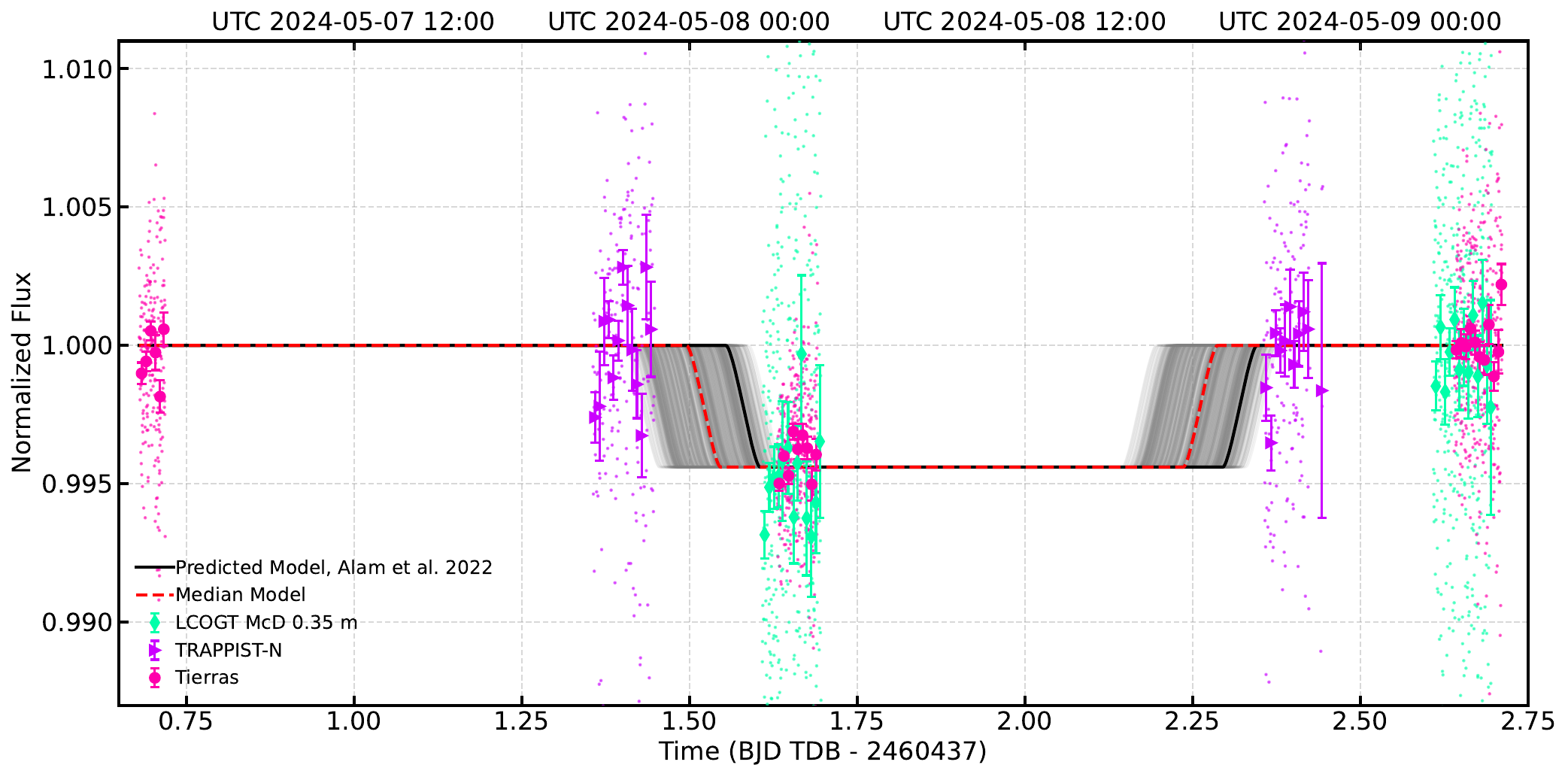} 
    \caption{The light curve for the transit of HIP~41378~f. The plot shows unbinned photometry (small, semi-transparent points) and 10~minute binned photometry (large symbols) from multi-night data sets. The data are color-coded by telescope: \textit{Tierras} (pink circles), TRAPPIST-North (violet flags), and LCOGT McD 0.35\,m (teal diamonds). The predicted transit model from \citetalias{Alam2022} ($T_{C,6} = 2460438.95 \pm 0.02$ BJD) is shown as a solid black line. Our median model is shown as a dashed red line \bedit{($T_{C,6} = 2460438.891 \pm 0.052$).} Gray lines represent 500 random model iterations drawn from the posterior density distribution shown in Figure~\ref{fig:histogram}. All model iterations assume $P = 542.07975$ days, $R_{\rm P}/R_\star = 0.0663$, $a/R_\star = 231.417$, $i = 89.971^{\circ}$, and $u = 0.0$ \citep{Santerne2019}. UTC dates corresponding to the observations are displayed at the top of the figure. \bedit{The LCOGT McD 1.0\,m datasets are not shown; see Section~\ref{sec:ntn_stable} and Figure~\ref{fig:ref_stars} for exclusion criteria.}} 
    \label{fig:lcfit} 
\end{figure*}

\begin{figure*}
\centering
\includegraphics[width=\linewidth]{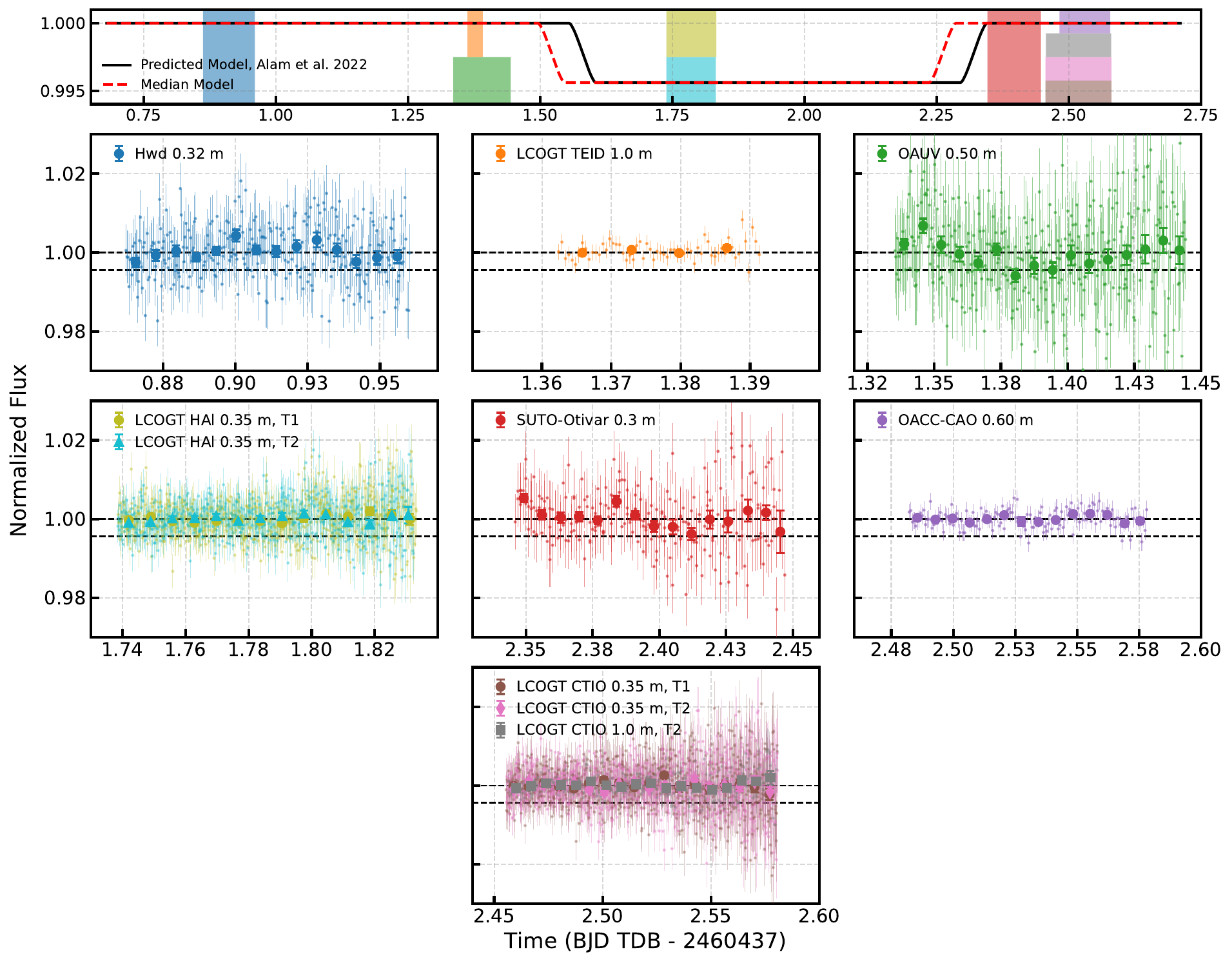}
\caption{\bedit{Top panel: Temporal coverage overview for all single-night observations during the 2024 May 7–9 campaign. Colored rectangles indicate the observation window for each telescope, vertically offset to show overlapping observations. The black solid line shows the predicted transit model from \citetalias{Alam2022}, and the red dashed line shows our fitted median model (Section \ref{sec:lcfit}). Bottom panels: Photometry from each single-night facility in the campaign, shown chronologically from top-left to bottom-center. Although these datasets span only a single night such that we cannot not use them in the light curve fit, the flatness of each time series provides visual confirmation that no transit ingress or egress is present. All data sets are median-normalized and detrended against airmass. We show unbinned fluxes (small, semi-transparent points) and 10~minute binned fluxes (large symbols), color-coded by telescope: Hwd (blue circles), LCOGT TEID (orange circles), OAUV (green circles), SUTO (red circles), OACC-CAO (purple circles), LCOGT CTIO 0.35\,m T1 (brown circles), LCOGT CTIO 0.35\,m T2 (pink diamonds), LCOGT CTIO 1.0\,m T2 (gray squares), LCOGT HAl 0.35\,m T1 (light blue circles) and LCOGT HAl 0.35\,m T2 (yellow triangles). The dashed black lines indicate the normalization baseline (unity) and the expected transit depth of HIP~41378~f. The flux dip observed in the OAUV data is significantly shorter than the expected transit duration.}}
\label{fig:singlenight_lcs}
\end{figure*}

\bedit{\subsection{Night-to-Night Photometric Stability Assessment}\label{sec:ntn_stable}}

\bedit{To assess whether our multi-night datasets possess sufficient photometric stability to detect the transit, we analyzed reference star flux variations. If significant flux variations are experienced only by HIP~41378, matching the expected 4.4~ppt transit depth of planet~f, this provides evidence both that the data have adequate stability and that we detected the transit.}

\begin{figure}
\centering
\includegraphics[width=.9\columnwidth]{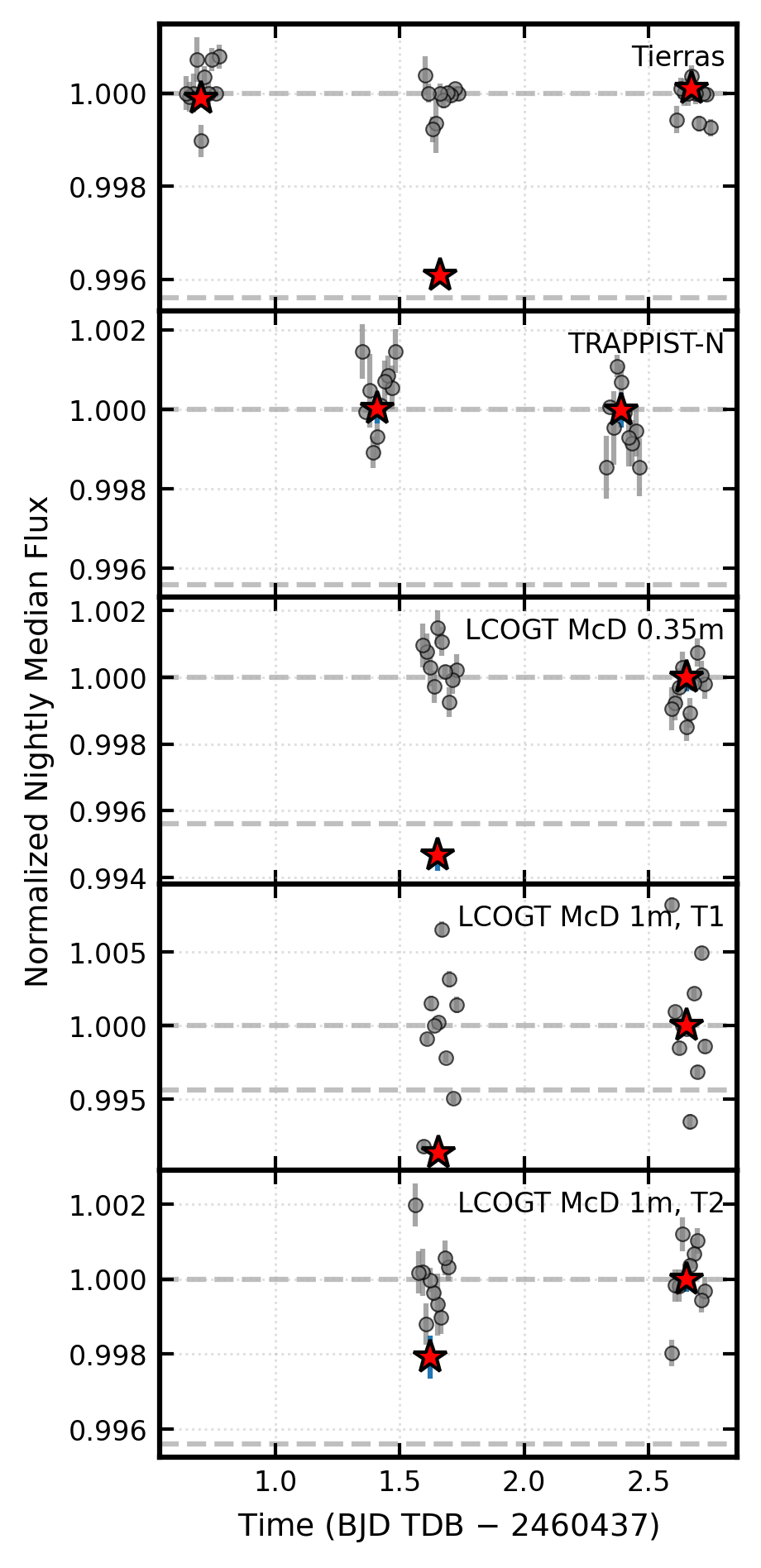}
\bedit{\caption {Normalized nightly median flux of the 10 highest-weighted reference stars (gray circles) compared to HIP~41378 (red star) for all multi-night facilities. Reference stars are shown in uniform gray; different stars with facility-specific weights are used for each telescope. Points are offset horizontally for clarity. The 10 stars represent 63\%, 98\%, 48\%, 61\%, and 51\% of the ELC weight for \textit{Tierras}, TRAPPIST-North, McD 0.35\,m, McD 1.0\,m T1, and McD 1.0\,m T2, respectively. Dashed lines mark unity and the 4.4~ppt expected transit depth. \textit{Tierras} and McD 0.35\,m show stable reference stars with HIP~41378 exhibiting 3.9~ppt and 5.3~ppt flux decrements on UTC 2024 May 8 (SNR $= 10\sigma$ and $4.8\sigma$). TRAPPIST-North is flat on both bracketing nights (1.12~ppt scatter, SNR $= 3.9\sigma$). McD 1.0\,m T1 exhibits 5.55~ppt reference star scatter exceeding the transit depth (SNR $= 0.8\sigma$). McD 1.0\,m T2's observing sequence was interrupted after 1.6 hours on the in-transit night (Table~\ref{tab:obs}), providing inadequate baseline characterization. Both McD 1.0\,m datasets are excluded from the transit fit.}\label{fig:ref_stars}}
\end{figure}

\bedit{For each multi-night dataset (\textit{Tierras}, TRAPPIST-North, LCOGT McD 0.35\,m, McD 1.0\,m T1, and McD 1.0\,m T2), we selected the 10 highest-weighted reference stars, representing 63\%, 98\%, 48\%, 61\%, and 51\% of each facility's ELC, respectively. For each reference star, we calculated the median nightly flux and computed flux ratios between consecutive observing nights. We quantified photometric stability via the night-to-night (NTN) scatter metric $\sigma_{\rm NTN} = \sigma(\text{flux ratios} - 1.0) / \sqrt{2}$, where the $\sqrt{2}$ correction accounts for independent measurements from two nights, each contributing photometric noise. We repeated this calculation for HIP~41378, normalizing its nightly median flux by the out-of-transit baseline (the median of two out-of-transit nights for \textit{Tierras} and TRAPPIST-North; the single out-of-transit night for the LCOGT McD datasets). The results are shown in Figure~\ref{fig:ref_stars}. \textit{Tierras} and TRAPPIST-North exhibit NTN scatter of 0.44 and 1.12~ppt, respectively. The LCOGT McD facilities show NTN scatter of 0.91~ppt (McD 0.35\,m), 5.55~ppt (McD 1.0\,m T1), and 1.22~ppt (McD 1.0\,m T2).}

\bedit{To assess detection capability, we calculated the signal-to-noise ratio (SNR) as the expected 4.4~ppt transit depth divided by the NTN scatter. This yields SNRs of 10.0$\sigma$ for \textit{Tierras}, 3.9$\sigma$ for TRAPPIST-North, and 4.8$\sigma$ for McD 0.35\,m, confirming that these three facilities possessed adequate precision to detect the transit. The measured flux decrements on UTC 2024 May 8 of 3.9~ppt (\textit{Tierras}) and 5.3~ppt (McD 0.35\,m) are consistent with the expected depth within their respective photometric uncertainties.}

\bedit{In contrast, McD 1.0\,m T1 exhibits NTN scatter (5.55~ppt) exceeding the expected transit depth (SNR $= 0.8\sigma$), indicating insufficient precision. McD 1.0\,m T2 is excluded for a different reason: its observing sequence was accidentally interrupted after only 53 exposures covering 1.6 hours on the in-transit night (Table~\ref{tab:obs}). While the facility's NTN scatter (1.22~ppt) appears adequate, this value is calculated from one complete out-of-transit night and one severely truncated observation. The brief baseline is inadequate for characterizing full-night photometric behavior and provides insufficient coverage to reliably distinguish astrophysical signals from systematic variations.}

\bedit{Figure~\ref{fig:ref_stars} visualizes these differences: reference stars in the \textit{Tierras}, TRAPPIST-North, and McD 0.35\,m datasets remain tightly clustered, with HIP~41378 exhibiting a clear flux decrement only on the in-transit night. In contrast, McD 1.0\,m T1 shows reference stars scattered by amounts exceeding the expected transit depth, while McD 1.0\,m T2 shows a measured transit depth commensurate with the reference star scatter. We therefore proceed with the transit analysis using only the three facilities (\textit{Tierras}, TRAPPIST-North, and McD 0.35\,m) that demonstrate both adequate photometric stability and sufficient observing baseline for transit detection.}

\bedit{\subsection{Transit Detection}}

\bedit{Figure~\ref{fig:ref_stars} demonstrates that HIP~41378 is a clear outlier in both the \textit{Tierras} and McD 0.35\,m datasets on UTC 2024 May 8: no reference star exhibits a comparable flux deviation on this night. After normalizing by the out-of-transit flux, we identified dips of around 4\,ppt and 5\,ppt in these datasets, matching the expected transit depth of HIP~41378~f (4.4\,ppt). The two TRAPPIST-North nights (UTC 2024 May 7 and 8) bracket the predicted transit window and are flat to within 0.74 ppt, ruling out early ingress or late egress. Together, these three datasets confirm the transit detection on UTC 2024 May 8.}

\bedit{Further evidence comes from single-night observations at the remaining facilities. Flat light curves from Hwd, LCOGT TEID, and OAUV rule out early ingress, while those from SUTO, OACC-CAO, and LCOGT CTIO rule out late egress. The two LCOGT HAl telescopes also exclude ingress or egress within the predicted window \citepalias{Alam2022}. These single-night data (Figure~\ref{fig:singlenight_lcs}) are not used in the light curve fit, as they cannot be reliably normalized without knowledge of each facility's flux baseline.}

\bedit{Combining photometric coverage from three UTC dates and ten geographically distinct observatories, we conclude with high confidence that the transit of HIP~41378~f occurred within the predicted window. The sub-mmag night-to-night photometric stability demonstrated by \textit{Tierras} \citep{Tamburo2025}, combined with the mmag stability of McD 0.35\,m and the bracketing observations from TRAPPIST-North, enabled detection of this shallow transit. This marks the fifth detected transit of the planet and only its second ground-based detection to date.}

\bedit{We prepared the \textit{Tierras}, TRAPPIST-North, and LCOGT McD 0.35\,m datasets for light curve modeling by normalizing each to its median out-of-transit flux and detrending each telescope-night against airmass \bedit{independently prior to transit fitting. This approach prevents degeneracy between the airmass correction and the transit center time, which is particularly important in the absence of a directly detected ingress or egress. We discuss the impact of this choice and compare it to simultaneous airmass detrending in Section~\ref{sec:lcfit}.} After 4$\sigma$ clipping (removing 6 exposures), 2317 photometric data points remained for the fit. In order to prevent underestimated photometric extraction errors from biasing our light curve fit results, we also scaled the normalized flux error bars of each individual telescope-night by $\alpha$: the ratio of the standard deviation of the normalized flux for that telescope-night to the median of its normalized flux errors (Table~\ref{tab:ratios}).}

\begin{table}[ht]
\caption{Normalized flux error scaling ratio, $\alpha$, per telescope-night used in the light curve fit.}
\begin{center}
\begin{tabular*}{\columnwidth}{@{\extracolsep{\fill}}lcc@{}}
\hline
Telescope & UTC Date\tablenotemark{\scriptsize{a}} & $\alpha$ \\
\hline
 \textit{Tierras} & 2024-05-07 & 1.00 \\
\textit{Tierras} & 2024-05-08 & 1.22 \\
\textit{Tierras} & 2024-05-09 & 1.33 \\
TRAPPIST-North & 2024-05-07 & 2.94 \\
TRAPPIST-North & 2024-05-08 & 3.10 \\
LCOGT McD 0.35\,m & 2024-05-08 & 2.67 \\
LCOGT McD 0.35\,m & 2024-05-09 & 2.40 \\
\hline
\end{tabular*}
\end{center}
\vspace{-15pt}
\tablenotetext{a}{The UTC Date listed is the start date of the observing sequence.}
\label{tab:ratios}
\end{table}

\section{Analysis}\label{sec:analysis}

\subsection{Light Curve Fit}\label{sec:lcfit}

We combined the multi-night data sets from \textit{Tierras}, TRAPPIST-North, and LCOGT McD \bedit{0.35\,m} to constrain the transit center time for the \bedit{2024 May 8} transit, $T_{C,6}$, by fitting a light curve to the data. We followed the notation from \cite{Bryant2021}, where $T_{C,N}$ refers to the time of the transit center for the transit epoch $N$. 

We used \texttt{batman} \citep{Kreidberg2015} to generate a light curve template with a period of $P = 542.07975$ days, a planet-to-star radius ratio of $R_{\rm P}/R_\star = 0.0663$, a scaled semi~major axis of $a/R_\star = 231.417$, and an inclination of $i = 89.971^{\circ}$ \citep{Santerne2019}. We set the initial mid-transit time of the template to $T_{C,6} = 2460438.95 \pm 0.02$ BJD TDB, following the prediction of \citetalias{Alam2022}. \bedit{We used a uniform (uninformative) prior on $T_{C,6}$, with the predicted value serving only as a starting point for the MCMC sampler. We adopted a linear limb-darkening law but fixed the coefficient to $u_1 = 0.0$, corresponding to a uniform stellar brightness profile. This choice was motivated by the presence of residual correlated noise in the \textit{Tierras} and McD 0.35\,m light curves. We tested the impact of this choice by performing fits with: (1) $u_1 = 0.0$ (frozen), (2) $u_1 = 0.3$ (frozen), (3) $u_1$ fitted with a uniform prior of 0.2–0.5 appropriate for Sun-like stars, and (4) $u_1$ fitted with a maximally permissive prior of 0.0–1.0. When $u_1$ was fixed to a more physical value ($u_1 = 0.3$; \citealt{Vanderburg2016}), the $T_{C,6}$ posterior median shifted earlier by $\sim$19 minutes but remained consistent within the 99.73\% credible interval. When $u_1$ was allowed to vary, the posteriors developed asymmetric tails and the fitted $u_1$ values were inconsistent between priors (0.305$\pm$0.08 vs 0.17$\pm$0.17), indicating that the transit model was partially absorbing red noise into the limb darkening parameter. Fixing $u_1 = 0.0$ yielded the most symmetric posterior and conservative uncertainty estimate on $T_{C,6}$, which is our primary scientific goal. Our initial light curve model is shown in Figure~\ref{fig:lcfit}.}

To fit the observational data and the light curve template, we set up a Markov Chain Monte Carlo Simulation (MCMC) using \texttt{edmcmc} \citep{TerBraak2006,Vanderburg2021}. The only free parameter of the fit is $T_{C,6}$. We ran the MCMC simulation with 500 walkers for a 3000 step burn-in phase, followed by a 7000 step chain. The chains were well mixed, as evidenced by \bedit{a \cite{Gelman1992} value of $\hat{R} - 1 = 0.0001$ for the free parameter.} 

\begin{figure}[h!]
    \centering
    \includegraphics[width=\linewidth]{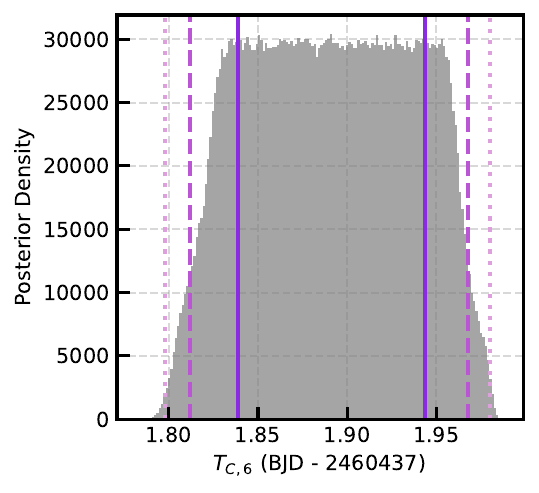} 
    \caption{The posterior density distribution of $T_{C,6}$ obtained from the MCMC simulation. Vertical lines indicate the limits of the 68.27\% (solid, dark purple), 95.45\% (dashed, medium purple), and 99.73\% (dotted, light purple) credibility intervals. The posterior is a uniform distribution since we did not catch HIP~41378~f during ingress or egress.}

    \label{fig:histogram} 
\end{figure}

We used a 7000-step MCMC chain to sample the posterior probability distribution of our free parameter, adopting the median as \bedit{our measured} value and the 16th–84th percentile range \bedit{(68\% credible interval) as the uncertainty}. We find \bedit{$T_{C,6} = 2460438.891 \pm 0.052$}. The \bedit{median} model is shown in Figure~\ref{fig:lcfit}. The absence of a clearly detected ingress or egress of HIP~41378~f results in a uniform posterior probability density for our constraint on $T_{C,6}$, as shown in Figure~\ref{fig:histogram}. We report our credibility intervals for $T_{C,6}$ in Table~\ref{tab:CIs}.

\bedit{We tested the robustness of our $T_{C,6}$ constraint to assumptions about per-telescope-night flux offsets. We repeated the MCMC fit while allowing one free offset per telescope-night, regularized by the empirically measured night-to-night photometric scatter values (Section~\ref{sec:ntn_stable}). The measured transit center time shifted by only 0.002 BJD ($\sim$3 minutes), from \bedit{$T_{C,6} = 2460438.891 \pm 0.052$} (fixed offsets) to $T_{C,6} = 2460438.893 \pm 0.053$ (fitted offsets), with identical confidence intervals and fit quality. This demonstrates that our $T_{C,6}$ constraint is insensitive to offset treatment. Given the negligible impact of fitting offsets, our empirical evidence that night-to-night baseline variations are small for these facilities (Section~\ref{sec:ntn_stable}), and the risk that floating offsets could absorb astrophysical signals in facilities without extensive night-to-night characterization \citep{Tamburo2025}, we adopt the fixed-offset model for our final result.}

\bedit{We detrended each telescope-night against airmass independently prior to transit fitting. To assess the impact of this choice, we also repeated the fit performing airmass detrending simultaneously with transit fitting, allowing the airmass slopes to vary freely as additional MCMC parameters. The simultaneous fit yields $T_{C,6} = 2460438.879$~BJD~TDB, approximately 17 minutes earlier than our adopted value, with modestly wider confidence intervals ($\pm1.43$~hours at 68\% versus $\pm1.26$~hours). The simultaneously fitted slopes systematically under-compensate for the airmass trends visible in the undetrended photometry, indicating that the MCMC is instead absorbing these systematics into the transit model. Specifically, the under-compensated airmass correction leaves systematically suppressed flux at high airmass (most notably late in the TRAPPIST-North 2024-05-08 dataset), which the model interprets as a potential early ingress relative to the \citetalias{Alam2022} prediction, shifting $T_{C,6}$ earlier. We therefore adopt independent detrending, which removes airmass systematics prior to fitting and prevents them from being misinterpreted as astrophysical signals. While overly aggressive detrending could in principle remove real transit signals, our analysis indicates the opposite risk is more significant here: simultaneous detrending can bias $T_{C,6}$ toward earlier values by creating a spurious ingress feature.}

\begin{table}[htbp]
    \centering
    \caption{Confidence intervals and error bars for $T_{C,6}$, the sixth transit epoch of HIP~41378~f, derived from the posterior distribution. The median value is \bedit{$2460438.891$} BJD TDB.}
    \begin{tabular}{cccc}
        \hline
        \hline
        Percentile & Confidence Interval & \multicolumn{2}{c}{Error Bar} \\
                  &                       & Lower       & Upper           \\
        \hline
        68.27\% & $2460438.839 - 2460438.943$ & $-0.052$ & $+0.052$ \\
        95.45\% & $2460438.812 - 2460438.968$ & $-0.079$ & $+0.077$ \\
        99.73\% & $2460438.798 - 2460438.980$ & $-0.093$ & $+0.089$ \\
        \hline
    \end{tabular}
    \label{tab:CIs}
\end{table}

\subsection{The period of HIP~41378~d}\label{sec:prettv} 
Previous TTV analyses of HIP~41378~f have typically assumed that its dominant perturber is planet~e, presumed to reside near a 3:2 MMR with planet~f \bedit{based on the RV-derived period of $369\pm10$ days from \citet{Santerne2019}}. However, only a single transit of planet~e has been observed, \bedit{and this RV estimate assumed planet d's period was fixed at 278 days. The transit-derived constraint on planet e's period is much broader ($P_e = 260^{+160}_{-60}$ days; \citealt{Lund2019}), and its radial-velocity signal lies near the instrumental precision limit, making it susceptible to systematics \citep{Santerne2019}.} In this section, we use \TESS\ Sector~88 data \citep{Ricker2015} to examine an alternative possibility: that HIP~41378~d may instead be the primary source of the observed TTVs of planet~f.

There is ongoing tension in the literature regarding the orbital period of HIP~41378~d. \citet{Grouffal2022} reported a partial RM detection consistent with a 278-day period, while \citet{Sulis2024} found no evidence of a transit near this period in \CHEOPS\ photometry. However, they showed that a TTV with an amplitude $>22.4$ hours could shift the transit outside the \CHEOPS\ observation window. The remaining viable orbital period aliases for planet~d are 101, 278, 371, and 1113~days.

\begin{figure}[h!]
    \centering
    \includegraphics[width=\linewidth]{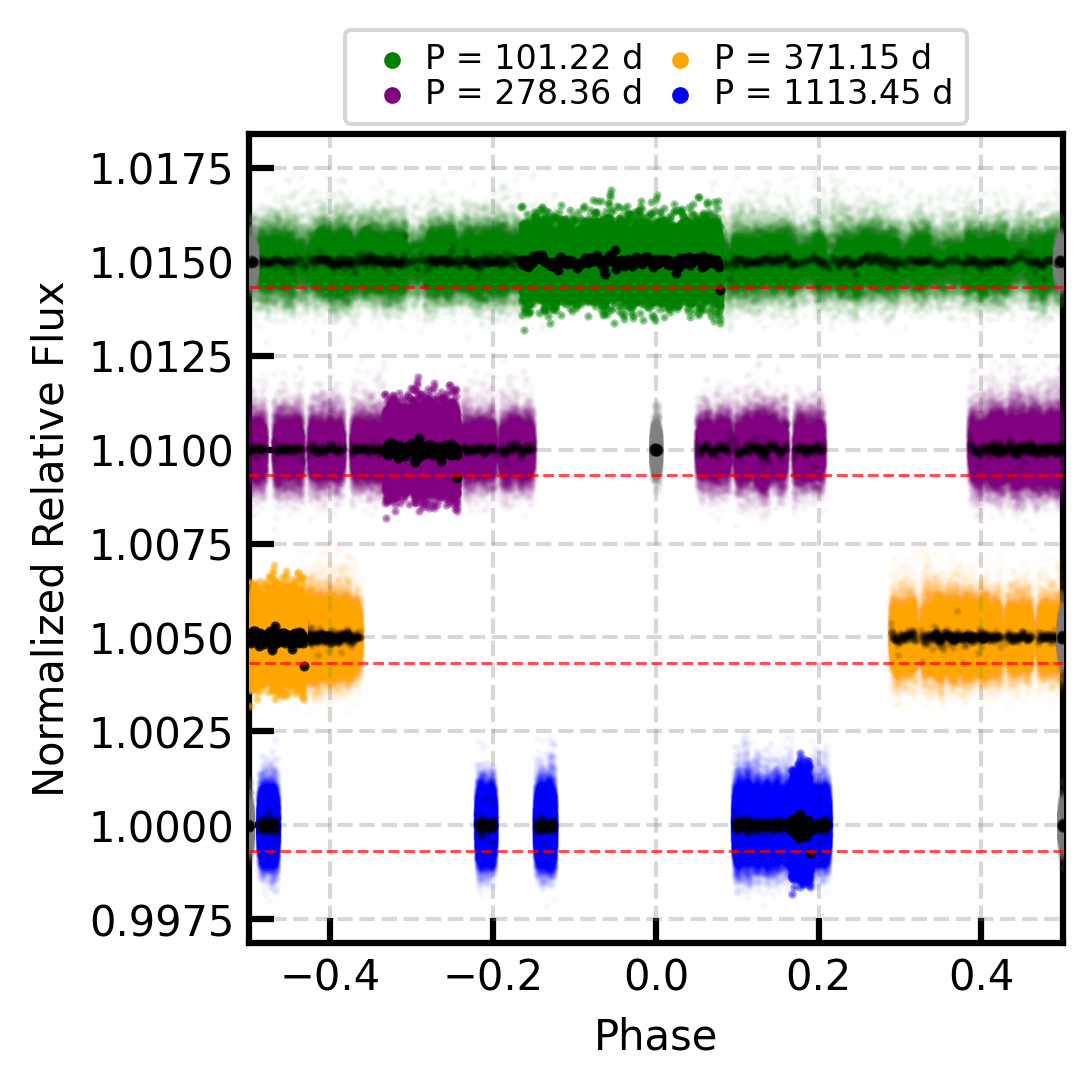}
    \caption{Building on the work from \cite{Sulis2024}, we plot all \TESS\ sectors that have targeted HIP~41378 (7, 34, 44, 45, 46, 61, 72, and 88) binned to 3 hours and phase-folded to the four possible orbital periods for HIP~41378~d ($P=101.2224$ d in green, $P=278.3616$ d in purple, $P=371.1488$ d in yellow, and $P=1113.4465$ d in blue). \bedit{Unbinned data are shown as small semi-transparent points, with \TESS\ Sector 88 data (higher opacity) highlighting the new observations that enable us to rule out the 101-day solution. Binned data are shown as larger black points. The red dashed lines indicate the expected transit depth of HIP~41378~d (0.67~ppt, \citealt{Becker2019}).} The 101-day solution is inconsistent with the \TESS\ data \bedit{at 7.2$\sigma$ confidence. Gray points show 38-sec \CHEOPS\ data from \cite{Sulis2024}.}}
    \label{fig:tess_88}
\end{figure}

HIP~41378 was observed by \TESS\ from 2025~January~14 to February~11 during Sector~88. We downloaded the corresponding 2-minute cadence data from MAST \bedit{\citep{10.17909/t9-nmc8-f686}} and combined it with all prior \TESS\ observations of this target (Sectors~7, 34, 44, 45, 46, 61, and 72). In Figure~\ref{fig:tess_88}, we show the light curve phase-folded to each of the four remaining period solutions for planet~d. The inclusion of Sector 88 data allows us to confidently rule out the 101-day solution. \bedit{We quantified the detection significance by calculating the photometric scatter in the 3-hour binned, phase-folded \TESS\ light curve for each period solution. The overall scatter is 0.09 ppt across all eight \TESS\ sectors. Given the expected transit depth of HIP 41378 d of 0.67 ppt ($R_p/R_\star = 0.0259 \pm 0.0015$, \citealt{Becker2019}), a transit would be detectable at 7.2$\sigma$ confidence. As shown in Figure~\ref{fig:tess_88}, no such transit feature is observed at phase zero for the 101-day solution, ruling out this period alias.}

\bedit{The minimum eccentricities for the remaining period solutions, derived from the observed transit duration constraint ($t_{\rm dur} \approx 12.5$ h; \citealt{Sulis2024}), are $e_{\rm min} \approx 0.14$ for the 278-day period, $e_{\rm min} \approx 0.21$ for the 371-day period, and $e_{\rm min} \approx 0.52$ for the 1113-day period. While these represent lower limits on the true eccentricities, dynamical stability arguments favor configurations closer to these minimum values in massive multi-planet systems \citep[e.g.,][]{Kipping2010}.}

The 1113-day period solution implies a minimum eccentricity of \( e \approx 0.52 \) to reproduce the observed transit duration \citep{Sulis2024}. At this eccentricity, the periastron distance of planet~d \((a_{\rm p} = a(1 - e) \approx 1.06~\mathrm{AU})\) lies well within the orbit of planet~f \((a = 1.367~\mathrm{AU})\), which is consistent with a near-circular orbit. Such an orbit-crossing configuration raises the possibility of dynamical instability unless protected by resonance, and may therefore disfavor the 1113-day solution.

\bedit{If TTVs are not invoked to explain the \CHEOPS\ non-detection near 278 days}, and if the 1113-day solution is dynamically disfavored, the remaining viable solution would correspond to an orbital period of approximately 371 days for planet~d, placing it near a 3:2 MMR with HIP~41378~f. \bedit{This is particularly notable given the $369\pm10$ day RV-derived period for planet~e discussed above.} If planet~d were to occupy this resonance, then the dynamical role and location of planet~e would remain uncertain. Table~\ref{tab:planetd_history} summarizes the current and conflicting constraints on the orbital period of HIP~41378~d.

\begin{table}[h!]
\caption{Summary of the remaining orbital period solutions for HIP~41378~d, building on the work from \cite{Sulis2024}.}
\begin{center}
\begin{tabular}{lcccc}
    \hline
    \hline
    Orbital period [d] & \TESS\ & \CHEOPS\ & RM & $e_{\text{min}}$ \\
    \hline
    $101.2224 \pm 0.0003$ & \xmark & \cmark & \xmark & $\sim$0.15 \\
    $278.3616 \pm 0.0009$ & \cmark & \xmark \bedit{\tablenotemark{\scriptsize a}} & \cmark & $\sim$0.14 \\
    $371.1488 \pm 0.0011$ & \cmark & \cmark & \xmark & $\sim$0.21 \\
    $1113.4465 \pm 0.0034$ & \cmark & \cmark & \xmark & $\sim$0.52 \\
    \hline
\end{tabular}
\end{center}
\tablecomments{We follow the convention established by \citet{Sulis2024}, where orbital periods compatible with the absence of a transit in \TESS\ or \CHEOPS\ observations are indicated by a (\cmark), while incompatible periods are marked with a (\xmark). The orbital period values are sourced from \citet{Becker2019}, and eccentricity values are from \citet{Sulis2024}. The RM column shows the results reported by \citet{Grouffal2022}.}
\tablenotetext{a}{\bedit{The \CHEOPS\ non-detection at $P = 278.36$ d could be explained by transit timing variations with amplitude $>22.4$ h \citep{Sulis2024}, though this scenario is in tension with the lack of detected TTVs over the 2.3-year baseline of RM observations presented in \citet{Grouffal2022} (see Section~\ref{sec:discussion}).}}
\label{tab:planetd_history}
\end{table}

\subsection{Transit Timing Variation Fit }\label{sec:ttvfit}

In this section, we update the TTV analysis for HIP~41378~f using our newly measured transit time, following the methods of \citet{Bryant2021} and \citetalias{Alam2022}. Similarly to \citet{Bryant2021}, we model the transit center time for the $N^{\rm th}$ epoch of HIP~41378~f as:

\begin{equation}\label{eq:ttv}
    T_{C,N} = T_0 + P_{\rm f} N + |V_{\rm f}|\sin\Big( 2\pi j \Delta N - \phi \Big)
\end{equation}

\noindent The first two terms describe a linear ephemeris, where \( T_0 \) is a reference transit time and \( P_{\rm f} \) is the orbital period of planet~f. The third term approximates the TTV induced by a near-resonant inner perturber, as derived by \citet{Lithwick2012}. Here, \( |V_{\rm f}| \) is the magnitude of the complex TTV amplitude and \( \phi \) is the TTV phase.

We estimate \( |V_{\rm f}| \) analytically following the first-order resonant formulation of \citet{Lithwick2012}, which parameterizes the amplitude as a function of the normalized distance to resonance, \( \Delta \):

\begin{equation}
    |V_{\rm f}(\Delta)| = P_{\rm f} \frac{\mu_{\rm in}}{\pi j \Delta} \left| -g(\Delta) + \frac{3}{2} \frac{f(\Delta)\, e_{\rm in} + g(\Delta)\, e_{\rm f}}{\Delta} \right|
\end{equation}

\noindent where

\begin{equation}
    \Delta = \left( \dfrac{P_{\rm f}}{P_{\rm in}} \cdot \dfrac{j - 1}{j} \right) - 1
\end{equation}

\noindent In these expressions, \( \mu_{\rm in} \) is the planet-to-star mass ratio of the inner planet, \( f(\Delta) \) and \( g(\Delta) \) are sums of Laplace coefficients with values near unity \citep[see Table~3 of][]{Lithwick2012}, and \( e_{\rm in} \), \( e_{\rm f} \) are the complex orbital eccentricities of the inner planet and HIP~41378~f, respectively.

We set up a second MCMC, also using \texttt{edmcmc} \citep{TerBraak2006,Vanderburg2021}, in order to fit all published transit times of HIP~41378~f \citep{Vanderburg2016,Becker2019,Bryant2021,Alam2022} and the RM-derived transit center time $T_{C,5} = 2459897.0199 \pm 0.0009$~BJD~TDB from \citet{Grouffal2025} to our transit time model (Eq.~\ref{eq:ttv}). The free parameters of the fit were $\Delta$, $P_{\rm f}$, $\mu_{\rm in}$, $\phi$, and $T_0$. The observational evidence (Section~\ref{sec:prettv}) suggests that either HIP~41378~e or d could be in a 3:2 MMR with HIP~41378~f. \bedit{Following \citet{Lithwick2012}, who note that their first-order formulation is valid for systems very close to resonance, we required $|\Delta| < 0.05$.} Given the uncertainty regarding the identity and mass of the inner companion of HIP~41378~f, we only required $\mu_{\rm in} > 0$. Finally, we set $-\pi < \phi < \pi$. \bedit{The first-order resonant TTV formalism has a well-known degeneracy between perturber mass and eccentricity \citep{Lithwick2012, Deck2015}, which cannot be broken without additional constraints (such as the ``chopping'' signal discussed below). We therefore fix $e_{\rm f} = 0.004$ \citep{Santerne2019} and test both $e_{\rm in} = e_{\rm d} = 0.21$ \citep{Sulis2024} and $e_{\rm in} = e_{\rm e} = 0.14$ \citep{Santerne2019}.}

Unlike previously measured transit epochs, our observational input for $T_{C,6}$ is a top-hat-shaped posterior rather than a Gaussian-distributed measurement (Figure~\ref{fig:histogram}). Consequently, instead of approximating the measurement as a Gaussian probability distribution, we approximate it as a uniform distribution bounded by the $99.73\%$ confidence interval. We implemented this in our code by rejecting all samples that predict a transit time outside the $99.73\%$ bounds of the posterior on $T_{C,6}$ (Table~\ref{tab:CIs}). This approach ensures that only TTV models consistent with the observed timing limits contribute to the final posterior distribution.

\begin{figure*}
    \centering
    \includegraphics[width=\linewidth]{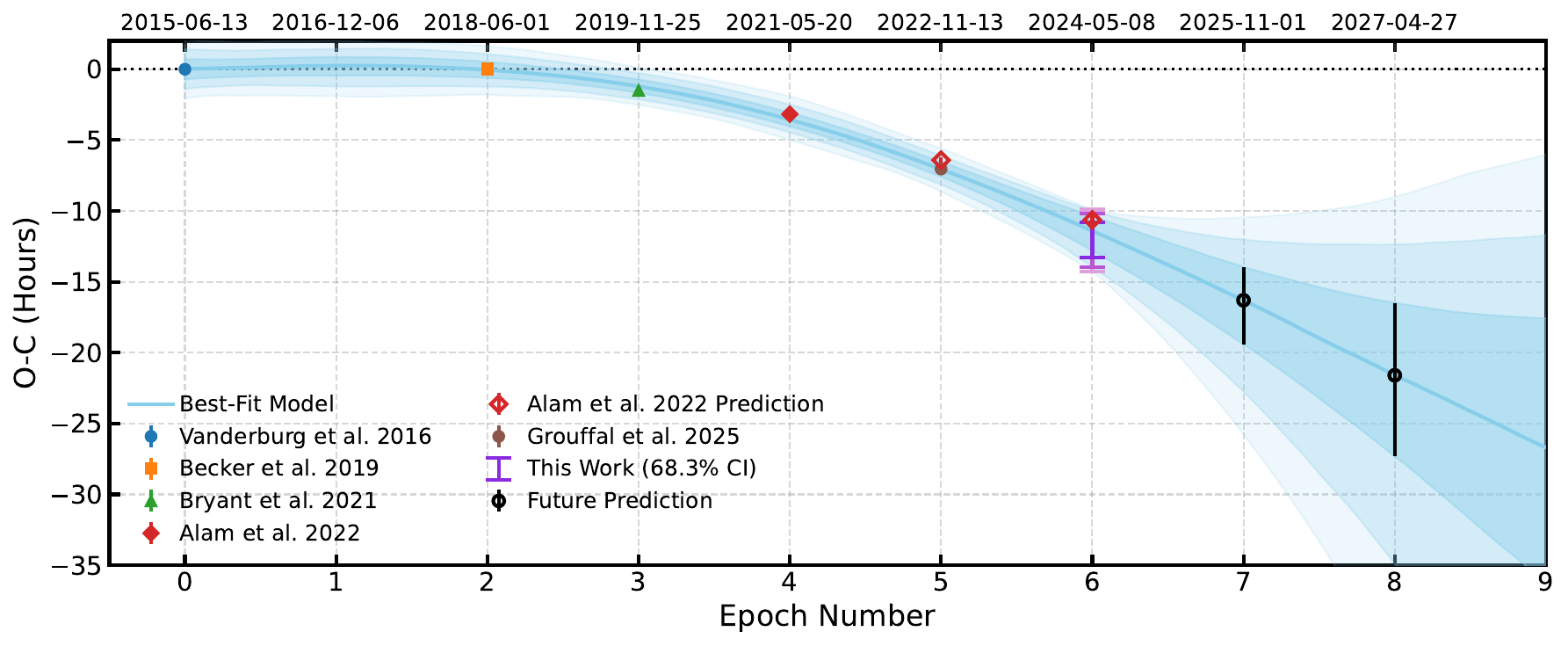} 
    \caption{Observed minus calculated (O--C) transit times as a function of epoch number for HIP~41378~f. Solid symbols denote previously published transits from \textit{K2} and NGTS analyses \citep{Vanderburg2016,Becker2019,Bryant2021}, as well as HST observations from \citetalias{Alam2022}, with associated error bars (smaller than the symbol size). The dotted black line represents a linear ephemeris using the period from \citet{Santerne2019} and the reference transit time from \citet{Vanderburg2016}. The solid blue curve shows our best-fit TTV model, with shaded bands indicating the 68.3\%, 95.4\%, and 99.7\% posterior predictive intervals. At epoch 6, our updated transit time is shown as a set of vertical purple error bars, with darker shades corresponding to the $68\%/95\%/99\%$ credible intervals in Figure~\ref{fig:histogram}. Open black circles at epochs 7 and 8 mark our predicted future transits, with statistical uncertainties shown as thick black error bars. The upper x-axis shows the corresponding UTC calendar dates.}

    \label{fig:ttvfit_d} 
\end{figure*}

\bedit{Our MCMC simulation used 100 walkers and $1.5 \times 10^7$ steps per walker, with a $5 \times 10^5$ step burn-in. The final chain was thinned by a factor of 100, resulting in $1.45 \times 10^7$ effectively independent samples used to characterize the posterior distributions of the free parameters.} Best-fit values were obtained by identifying the parameter set that maximized the log-likelihood. The results of the fit (\bedit{assuming $e_{\rm in} = e_{\rm d} = 0.21$) are shown in Figure~\ref{fig:ttvfit_d}. We note that the results for $e_{\rm in} = e_{\rm e} = 0.14$ are essentially identical.}

\bedit{The first-order formulation of \citet{Lithwick2012} captures the long-period, sinusoidal TTV signal generated by near-resonant interactions and is the appropriate model for this system given the limited number of observed transits, which cannot constrain a more complex dynamical model. However, by design this formalism does not account for the short-period ``chopping'' component arising from synodic conjunctions \citep{Deck2015}. To properly use the \citet{Lithwick2012} model, we must account for this missing signal as a systematic uncertainty. We used \texttt{TTVFaster} \citep{Agol2016} to estimate the expected chopping amplitude for HIP~41378~f, computing the full TTV signal and isolating the chopping component by excluding the dominant resonant harmonics. Assuming either planet~d ($P = 371$~d) or planet~e ($P = 369$~d) as the perturber, we find RMS chopping amplitudes of 0.27 and 0.72~hours, respectively. We adopt 0.72~hours as a conservative estimate of this systematic and inflate the published transit time measurement uncertainties by this amount (added in quadrature) before fitting. Table~\ref{tab:future_transits} lists our predicted transit center times for epochs 7 and 8.}

\begin{table}[ht]
\centering
\caption{Predicted future transits of HIP~41378~f. Statistical uncertainties represent the 68.3\% posterior predictive credible intervals.}
\setlength{\tabcolsep}{6pt} 
\begin{tabular}{lcr}
\hline
Epoch & $T_C$ (BJD TDB) & UTC Date \\
\hline
7 & $2460980.793 ^{+0.098}_{-0.129}$   & 2025 Nov 1 \\
8 & $2461522.653 ^{+0.213}_{-0.238}$   & 2027 Apr 27 \\
\hline
\end{tabular}
\label{tab:future_transits}
\end{table}

\section{Discussion and Conclusions}\label{sec:discussion}

\bedit{We observed a transit of HIP~41378~f between 2024~May~7 and 9 using data from ten geographically distributed observatories. Although ingress and egress were not detected in any single light curve, we identified consistent flux dips in the multi-night observations from \textit{Tierras} and LCOGT McD 0.35\,m that match the expected transit depth of HIP~41378~f. After normalizing by the apparent out-of-transit flux and reprocessing using the \textit{Tierras} pipeline, we measured flux decrements of 3.9 ppt and 5.3 ppt on UTC 2024 May 8, consistent with the expected 4.4 ppt transit depth to within each facility's night-to-night scatter (0.44 ppt for \textit{Tierras} and 0.91 ppt for McD 0.35\,m). The TRAPPIST-North data bracket the predicted transit window with a night-to-night scatter of 1.12 ppt, well below the transit depth, ruling out early ingress and late egress. Additional flat light curves from other sites further support this conclusion. Together, these datasets confirm that the planet was in transit during our observations and allow us to constrain the transit center time, even in the absence of a clearly observed ingress or egress. The median transit time is $2460438.891$ BJD TDB and the confidence intervals are listed in Table~\ref{tab:CIs}.}

\bedit{We attribute the poor photometric performance of the LCOGT McD 1.0\,m telescopes to the use of the Sloan $z_s$ filter ($\lambda_c \approx 950$~nm), which samples wavelengths heavily affected by precipitable water vapor (PWV) variations in Earth's atmosphere \citep{St07,Bl08,Be12}. In contrast, \textit{Tierras} employs a custom narrow-band filter specifically designed to minimize PWV errors ($\lambda_c = 863.5$~nm, FWHM = 40~nm; \citealt{GaMe20}), McD 0.35\,m uses the Sloan $i'$ filter ($\lambda_c \approx 760$~nm) where water absorption is reduced, and TRAPPIST-North uses a $B$ filter ($\lambda_c \approx 450$~nm) at blue wavelengths where water vapor lines are suppressed. Although the $i'$ filter would be preferable for the LCOGT 1.0\,m network, it is impractical when observing HIP 41378 owing to a 3\% duty cycle (1~sec exposure, 30~sec readout). This comparison demonstrates the critical importance of filter selection for achieving sub-mmag night-to-night photometric precision from the ground, particularly when detecting shallow transits of long-period planets.}

This marks only the second ground-based detection of a transit of HIP~41378~f, making it the longest period planet to have been observed transiting twice from the ground. This was an unconventional observing mode for many of the facilities involved in the campaign, providing valuable lessons. To improve the efficacy of ground-based longitudinal coverage observations for long-period, long-duration transit events in the future, it would be beneficial to optimize the overlap in the field of view between facilities, thereby increasing the availability of comparison stars common to all data sets. Furthermore, ensuring that at least one night of out-of-transit data is collected per facility, even a few nights before or after  the predicted transit window, could improve the transit center constraint by providing knowledge of each facility's baseline.  

The orbital period of HIP~41378~d remains uncertain. We rule out the 101-day solution using eight \TESS\ sectors. The 278-day solution supported by \citet{Grouffal2022}'s partial RM detection remains viable. \citet{Sulis2024} showed that this period is consistent with \CHEOPS\ data if the TTV amplitude exceeded 22.4 hours, which their dynamical model permits. However, this is in tension with the RM-based claim of no TTVs over a 2.3-year baseline, raising questions about the plausibility of such rapid dynamical evolution. The 1113-day solution appears dynamically unstable. Finally, the 371-day alternative conflicts with the reported 278-day RM detection and the 369-day period attributed to planet e \citep{Santerne2019}. Improved RM measurements and future transits of HIP~41378~d and e, although challenging to observe, will be critical to resolving this ambiguity.

An intriguing avenue for future work lies in jointly modeling the TTVs of planet~f and the nondetection of planet~d by \CHEOPS\ to better constrain the mass and orbital properties of planet~e. Although only a single transit of planet~e has been observed, its dynamical influence could be significant. If planets~d, e, and f lie near a 3:4:6 MMR, as suggested by \citet{Santerne2019}, then self-consistent dynamical fits that incorporate both the TTV signal and the \CHEOPS\ constraints may clarify the role of planet~e in shaping the observed timing variations of planet~f.

Similarly to \citetalias{Alam2022} and \citet{Bryant2021}, we modeled the TTV signal of HIP~41378~f using the first-order formulation proposed by \citet{Lithwick2012}, which is appropriate for planets near (but not in) MMR. \bedit{This formulation is appropriate for this system given the limited number of observed transits, but by design it neglects the short-period chopping signal associated with synodic conjunctions. As seen in Figure~\ref{fig:ttvfit_d}, the residuals between the published transit times and our model indicate that even accurate measurements will scatter around \citet{Lithwick2012} predictions due to this unmodeled signal. We therefore inflated the measurement uncertainties by 0.72~hours (added in quadrature) to account for the expected chopping amplitude (Section~\ref{sec:ttvfit}).}

We note that the TTVs of HIP~41378~f exhibit a predominantly quadratic trend, leading to significant covariance between the TTV period and the amplitude in our posterior distribution. If the TTV period is very long, this could indicate that the perturbing planet lies extremely close to resonance with HIP~41378~f. In such cases, the assumptions of the near-resonant approximation, and the interpretation of the amplitude $|V_{\rm f}|$, may begin to break down, as higher-order terms and resonant effects become important \citep{Deck2015}. While this likely affects only a portion of the posterior, future analyses using full dynamical models that incorporate both resonant and synodic contributions may yield more accurate inferences about the perturber's identity and orbital architecture.

The hour-wide transit time center window for HIP~41378~f predicted by \citetalias{Alam2022} falls within our constraint's $99.7\%$ confidence interval, but since the data cannot rule out earlier transit times (and therefore larger TTV values), we can only state that the transit arrived between 10 and 14.\bedit{4} hours earlier than expected from a linear ephemeris with $99.7\%$ confidence. We predict that the next transits of HIP~41378~f will occur on UTC 2025 November 1 and 2027 April 27 (Table~\ref{tab:future_transits}). 

Currently, there are no scheduled space-based observations of HIP~41378~f despite the wealth of information that could be gained from studying the planet's dynamical and chemical context. Observing such a long-duration transit with a precise space-based photometer such as \CHEOPS\ \citep{Benz2021} would overcome the longitudinal coverage challenges inherent to ground-based observations, allowing us to place tighter constraints on the TTV signal of HIP~41378~f and potentially the mass, period, eccentricity, and identity of its perturber.

As the only bright ($J < 10$), long-period ($P > 1$~yr) transiting exoplanet discovered to date, HIP~41378~f offers a unique opportunity for atmospheric characterization with \textit{JWST}. Although \citetalias{Alam2022} reported a flat transmission spectrum in the near-infrared using HST/WFC3, mid-infrared spectroscopy with \textit{JWST} could distinguish between (optically thin) photochemical hazes and (optically thick) circumplanetary rings as the primary cause of the planet's low density \citep{Aki2020, Lu2025}. The upcoming transits of HIP~41378~f in November 2025 and April 2027 present a rare opportunity to probe the dynamical and chemical properties of this intriguing system.


\begin{center}
    ACKNOWLEDGMENTS
\end{center}

JGM gratefully acknowledges support from the Heising-Simons Foundation and the Pappalardo family through the MIT Pappalardo Fellowship in Physics. The postdoctoral fellowship of KB is funded by F.R.S.-FNRS grant T.0109.20 and by the Francqui Foundation. \bedit{Funding for KB was provided by the European Union (ERC AdG SUBSTELLAR, GA 101054354).} MG and EJ are F.R.S.-FNRS Research Directors. APo acknowledges grant BK-2025. This work makes use of observations from the \textit{Tierras} Observatory, which is supported by the National Science Foundation under Award No. AST-2308043. \textit{Tierras} is located within the Fred Lawrence Whipple Observatory; we thank all the staff there who help maintain this facility. The research leading to these results has received funding from  the ARC grant for Concerted Research Actions, financed by the Wallonia-Brussels Federation. TRAPPIST is funded by the Belgian Fund for Scientific Research (Fond National de la Recherche Scientifique, FNRS) under the grant PDR T.0120.21. TRAPPIST-North is a project funded by the University of Liege (Belgium), in collaboration with Cadi Ayyad University of Marrakech (Morocco). This work makes use of observations from the LCOGT network. Part of the LCOGT telescope time was granted by NOIRLab through the Mid-Scale Innovations Program (MSIP). MSIP is funded by NSF. This paper is based on observations made with the Las Cumbres Observatory’s education network telescopes that were upgraded through generous support from the Gordon and Betty Moore Foundation. Funding for the \TESS\ mission is provided by NASA's Science Mission Directorate. KAC and CNW acknowledge support from the \TESS\ mission via subaward s3449 from MIT. This paper includes data collected by the \TESS\ mission, which are publicly available from the Mikulski Archive for Space Telescopes (MAST). This research has made use of the Exoplanet Follow-up Observation Program \bedit{(ExoFOP; \citealt{10.26134/ExoFOP5})} website, which is operated by the California Institute of Technology, under contract with the National Aeronautics and Space Administration under the Exoplanet Exploration Program. We acknowledge financial support from the Agencia Estatal de Investigaci\'on of the Ministerio de Ciencia e Innovaci\'on MCIN/AEI/10.13039/501100011033 and the ERDF “A way of making Europe” through project PID2021-125627OB-C32, and from the Centre of Excellence “Severo Ochoa” award to the Instituto de Astrof\'isica de Canarias. We are grateful to the Unistellar Telescope Network, ARTEMIS-SPECULOOS, and the Van Vleck Observatory for attempting observations as part of this campaign. 


\facilities{\textit{Tierras}, TRAPPIST-North, LCOGT, \TESS, Mikulski Archive for Space Telescopes (MAST), GAIA}

\software{\bedit{\textit{Astropy} \citep{astropy:2013,astropy:2018,astropy:2022}}, \bedit{\textit{numpy} \citep{numpy}}, \textit{pandas} \citep{pandas}, \textit{scipy} \citep{scipy}, \textit{matplotlib} \citep{matplotlib}, \textit{batman} \citep{Kreidberg2015}, AstroImageJ \citep{Collins2017}, \textit{lightkurve} \citep{lightkurve}, \textit{edmcmc} \citep{Vanderburg2021}}, \textit{TTVFaster} \citep{Agol2016,Agol2016soft}

\bibliography{hip41378fbib}{}
\bibliographystyle{aasjournal}



\end{document}